\documentclass[%
    aps,
    prd,
    twocolumn,
    reprint,
    10pt,
    nofootinbib,
    superscriptaddress,
    longbibliography,
    amsfonts,
    amssymb,
    amsmath,
    preprintnumbers,
    noshowpacs,
    noshowkeyws,
    tightenlines,
    floats,
    notitlepage,
    final,
    letterpaper,
    oneside
    ]{revtex4-1}
\pdfoutput=1
\usepackage[utf8]{inputenc}
\usepackage[T1]{fontenc}
\usepackage[british]{babel}
\usepackage[british,cleanlook,printdayon]{isodate}
\usepackage[Symbolsmallscale]{upgreek}
\usepackage{mathtools}
\usepackage[protrusion,tracking,kerning,spacing,final,babel]{microtype}
\usepackage{physics}
\usepackage[dvipsnames]{xcolor}
\usepackage{graphicx}
\usepackage[final]{hyperref}
\usepackage[normalem]{ulem}
\usepackage{subfigure}
\hypersetup{%
    colorlinks=true,
    citecolor=MidnightBlue,
    linkcolor=MidnightBlue,
    urlcolor=MidnightBlue
    }%
\usepackage{cleveref}
\usepackage{nth}

\newcommand*{\rme}{\mathrm{e}}
\newcommand*{\rmi}{\mathrm{i}}
\newcommand*{\obs}{\mathrm{obs}}
\newcommand*{\orb}{\mathrm{orb}}
\newcommand*{\sol}{\mathrm{sol}}
\renewcommand*{\var}{\mathrm{Var}}
\newcommand*{\cov}{\mathrm{Cov}}
\newcommand*{\cald}{\mathcal{D}}
\newcommand*{\calt}{\mathcal{T}}
\newcommand*{\call}{\mathcal{L}}
\newcommand*{\caln}{\mathcal{N}}
\newcommand*{\calp}{\mathcal{P}}
\newcommand*{\hatd}{\hat{\mathcal{D}}}
\newcommand*{\hatp}{\hat{\mathcal{P}}}

\begin{document}
\title{Targeted search for the kinematic dipole of the gravitational-wave background}

\author{Adrian Ka-Wai Chung}
\email{ka-wai.chung@ligo.org}
\affiliation{Theoretical Particle Physics and Cosmology Group, Physics Department, King's College London, University of London, Strand, London WC2R 2LS, UK}

\author{Alexander C. Jenkins}
\email{alex.jenkins@ucl.ac.uk}
\affiliation{Department of Physics and Astronomy, University College London, London WC1E 6BT, UK}

\author{Joseph D. Romano}
\email{joseph.d.romano@ttu.edu}
\affiliation{Department of Physics and Astronomy, Texas Tech University, Box 41051, Lubbock, TX 79409-1051, USA}

\author{Mairi Sakellariadou}
\email{mairi.sakellariadou@kcl.ac.uk}
\affiliation{Theoretical Particle Physics and Cosmology Group, Physics Department, King's College London, University of London, Strand, London WC2R 2LS, UK}

\date{\today}
\preprint{KCL-PH-TH/2022-38}

\begin{abstract}
    There is growing interest in using current and future gravitational-wave interferometers to search for anisotropies in the gravitational-wave background.
    One guaranteed anisotropic signal is the kinematic dipole induced by our peculiar motion with respect to the cosmic rest frame, as measured in other full-sky observables such as the cosmic microwave background.
    Our prior knowledge of the amplitude and direction of this dipole is not explicitly accounted for in existing searches by LIGO/Virgo/KAGRA, but could provide crucial information to help disentangle the sources which contribute to the gravitational-wave background.
    Here we develop a targeted search pipeline which uses this prior knowledge to enable unbiased and minimum-variance inference of the dipole magnitude.
    Our search generalises existing methods to allow for a time-dependent signal model, which captures the annual modulation of the dipole due to the Earth's orbit.
    We validate our pipeline on mock data, demonstrating that neglecting this time dependence can bias the inferred dipole by as much as $\order{10\%}$.
    We then run our analysis on the full LIGO/Virgo O1+O2+O3 dataset, obtaining upper limits on the dipole amplitude that are consistent with existing anisotropic search results.
\end{abstract}

\maketitle

\begin{figure*}[t!]
    \centering
    \includegraphics[width=0.75\textwidth]{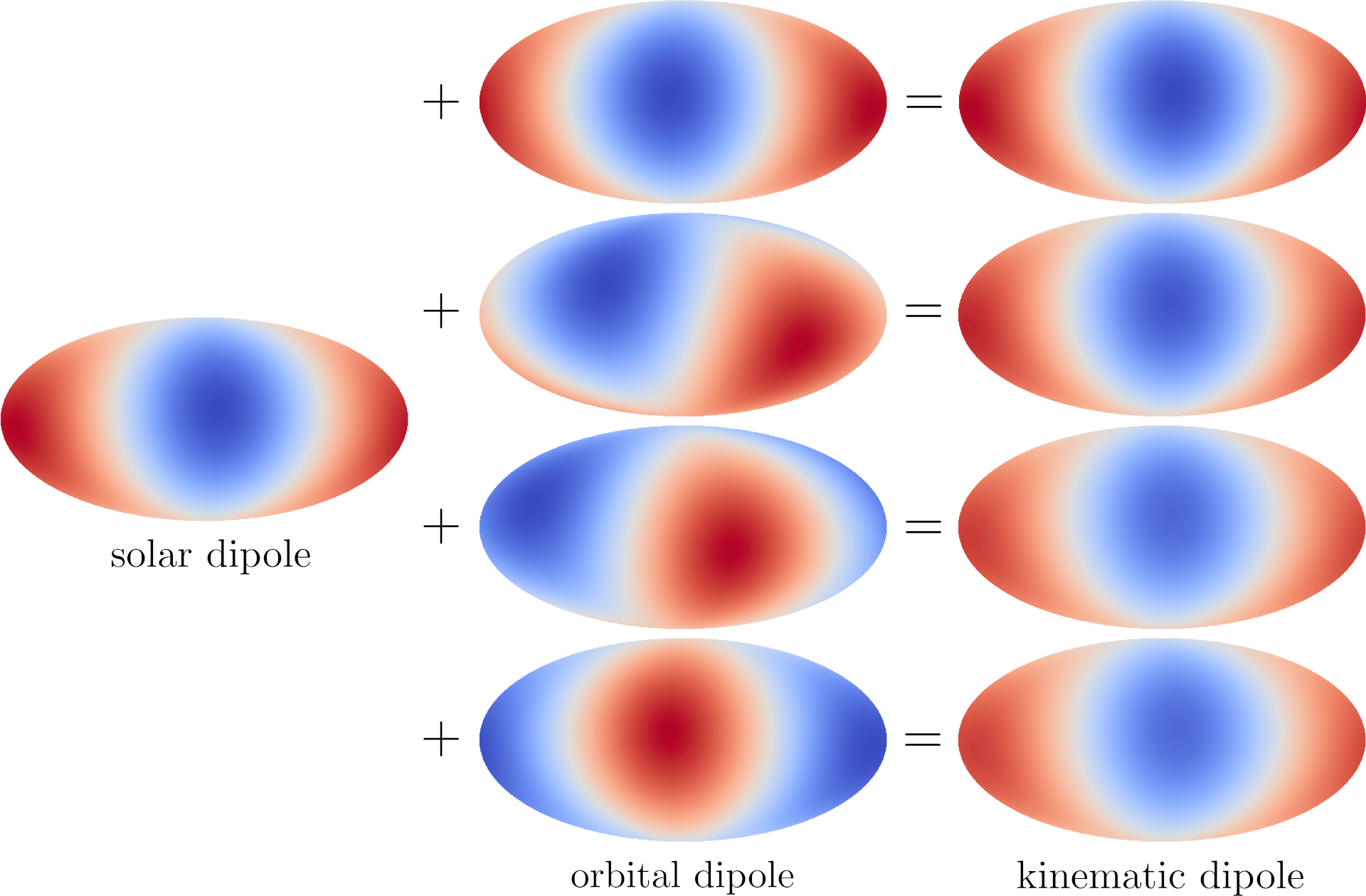}
    \caption{The kinematic dipole is a superposition between a (time-independent) \emph{solar dipole} and a (time-dependent) \emph{orbital dipole}.
    Here we show the orbital dipole at 0, 2, 4, and 6 months from the time at which the Earth is at perihelion (reading from top to bottom), with the amplitude of the total dipole becoming progressively smaller over this time.
    The resulting $\order{10\%}$ variation is difficult to distinguish by eye here, but has a significant impact on the search, as demonstrated in figure~\ref{fig:new_injections}.}
    \label{fig:maps}
\end{figure*}

\section{Introduction}

The direct detection of gravitational waves (GWs) from compact binary coalescences by the LIGO/Virgo/KAGRA Collaboration (LVK) has initiated the era of GW astronomy~\cite{LIGOScientific:2018mvr,LIGOScientific:2020ibl,LIGOScientific:2021usb,LIGOScientific:2021djp}.
As well as transient, individually-resolvable signals such as these, one also expects a gravitational-wave background (GWB) due to the superposition of GWs produced by many weak, independent and unresolved sources of cosmological or astrophysical origin~\cite{Allen:1996vm,Maggiore:1999vm,Romano:2016dpx,Christensen:2018iqi,Renzini:2022alw}.
Once detected, this background will provide interesting astrophysical information about the formation of black holes and neutron stars throughout cosmic time~\cite{Regimbau:2011rp,Martinovic:2021fzj}, and will potentially shed light on early universe cosmology and particle physics beyond the Standard Model~\cite{Caprini:2018mtu,LIGOScientific:2021nrg,Romero:2021kby,Martinovic:2021hzy,Romero-Rodriguez:2021aws,Sakellariadou:2022tcm}.

In order to extract as much of this information as possible from the GWB, we must study not only its mean intensity, but also the fluctuations in this intensity across the sky---the \emph{anisotropies} in the GWB.
There has been much recent interest in studying the ``intrinsic'' anisotropies due to statistical clustering of GW sources, particularly for the astrophysical GWB from coalescing compact binaries~\cite{Cusin:2017fwz,Cusin:2017mjm,Jenkins:2018nty,Cusin:2018rsq,Geller:2018mwu,Jenkins:2018uac,Jenkins:2018kxc,Bartolo:2019zvb,Bertacca:2019fnt,Bartolo:2019oiq,Bartolo:2019yeu,Bellomo:2021mer,LISACosmologyWorkingGroup:2022kbp}.
However, as first pointed out in~\cite{Jenkins:2019uzp,Jenkins:2019nks}, the relatively small number of compact binaries in the observable Universe means that measurements of these anisotropies suffer from very high levels of shot noise, limiting the astrophysical and cosmological information that can be extracted from them.

Aside from the intrinsic anisotropies, there are also \emph{extrinsic} or \emph{kinematic} anisotropies due to the peculiar motion of our GW detectors with respect to the rest frame of the GWB~\cite{Jenkins:2018nty,Jenkins:2018uac,Cusin:2022cbb,Jenkins:2022chv,DallArmi:2022wnq}.
Most notable of these is the Doppler enhancement of the GW intensity we observe from sources we are moving toward, and the corresponding reduction in intensity from sources we are receding from.
This generates a \emph{kinematic dipole} in the GWB.
This effect is well-known, and has been measured with high precision in other cosmological observables, particularly the cosmic microwave background (CMB)~\cite{Planck:2018nkj}.
As we discuss below, a joint measurement of this dipole and of the monopole could help us unravel the physics of the sources that contribute to the GWB, as well as potentially shedding light on the existing tension between the CMB dipole and the dipole observed in source counts of quasars and radio galaxies~\cite{Secrest:2020has,Dalang:2021ruy,Secrest:2022uvx} (see however~\cite{Darling:2022jxt}).

We present here a targeted search method to measure the GWB kinematic dipole.
In \cref{sec:KD-model} we construct a signal model of the dipole by considering both the (constant) motion of the solar system barycentre with respect to the cosmic rest frame, and the (time-dependent) orbital motion of the Earth around this barycentre, as illustrated in figure~\ref{fig:maps}.
In \cref{sec:Method} we develop a maximum-likelihood estimator for the mean dipole amplitude, extending the existing data-analysis search formalism~\cite{Romano:2016dpx,KAGRA:2021kbb} to allow for a time-dependent signal.
In \cref{sec:PE} we validate our formalism on mock data, before running our analysis on the first three observing runs (O1, O2, O3) of the Advanced LIGO and Advanced Virgo interferometers.
We summarise our conclusions in \cref{sec:conclusion}.

We work in relativistic units, $c=1$, throughout.
Following~\cite{LIGOScientific:2016jlg,LIGOScientific:2019vic,KAGRA:2021kbb,LIGOScientific:2016nwa,LIGOScientific:2019gaw,KAGRA:2021mth}, we assume the \emph{Planck} 2015 value of the Hubble constant, $H_0=67.9\,\mathrm{km~s^{-1}~Mpc^{-1}}$~\cite{Planck:2015fie}.

\section{Signal model}
\label{sec:KD-model}

\subsection{The monopole}
\label{sec:monopole}

We characterise the isotropic (monopole) component of the GWB in terms of its energy density spectrum (measured in units of the cosmological critical energy density $\rho_\mathrm{c}\equiv3H_0^2/(8\uppi G)$),
    \begin{equation}
        \Omega(f)\equiv\frac{1}{\rho_\mathrm{c}}\dv{\rho_\mathrm{gw}}{(\ln f)}.
    \end{equation}
We calculate this spectrum in terms of quantities describing the GW sources by writing \cite{Phinney:2001di}
    \begin{equation}
    \label{eq:monopole}
        \Omega(f)=\frac{1}{\rho_\mathrm{c}}\int_0^\infty\frac{\dd{z}}{(1+z)^2}\int\dd{\vb*\vartheta}\frac{\mathcal{R}(z,\vb*\vartheta)}{H(z)}\dv{E(\vb*\vartheta)}{(\ln f_\mathrm{s})},
    \end{equation}
    where $H(z)$ is the Hubble rate, $\vb*\vartheta$ is a set of parameters characterising the sources, $\mathcal{R}$ is the comoving source rate density, and $\dv*{E}{(\ln f_\mathrm{s})}$ is the spectrum of the radiated GW energy in terms of the source-frame frequency $f_\mathrm{s}=(1+z) f$.

For simplicity, we assume that this spectrum is well-approximated by a power law throughout our sensitive frequency band,\footnote{%
    Note that this power-law assumption is \emph{not} essential to our search, and can easily be replaced with a generic spectral ansatz.}
    \begin{equation}
        \Omega(f)=\Omega_\alpha(f/f_\mathrm{ref})^\alpha,
    \end{equation}
    with $\Omega_\alpha$ giving the GWB amplitude at some reference frequency $f_\mathrm{ref}$.
(Following~\cite{LIGOScientific:2016jlg, LIGOScientific:2019vic, KAGRA:2021kbb, LIGOScientific:2016nwa, LIGOScientific:2019gaw, KAGRA:2021mth}, we use $f_\mathrm{ref}=25\,\mathrm{Hz}$ throughout.)
LVK stochastic analyses typically focus on three $\alpha$ values of particular interest:
    \begin{itemize}
        \item $\alpha=0$, which corresponds to a scale-invariant background (i.e. $\Omega=\mathrm{constant}$).
        This is approximately what one expects for a GWB signal from Nambu-Goto cosmic string loops at high frequencies~\cite{LIGOScientific:2021nrg}, and is also characteristic of (single-field, slow-roll) inflationary models~\cite{Allen:1996vm,Maggiore:1999vm}.
        \item $\alpha=2/3$, which corresponds to a (slightly) blue-tilted spectrum.
        This is a robust prediction for the astrophysical GWB from inspiralling (population I or II) compact binaries~\cite{Regimbau:2011rp}.
        \item $\alpha=3$, which corresponds to a white-noise strain power spectral density $S_\mathrm{gw}(f)\propto\Omega(f)/f^3=\text{const}$.
        This has been studied as an approximation to the astrophysical GWB from core-collapse supernovae~\cite{Buonanno:2004tp}.
    \end{itemize}
In the following sections, we present results for each of these three spectral indices.

\subsection{The kinematic dipole}

The Earth's peculiar motion with respect to the GWB rest frame induces a small, time-dependent, anisotropic contribution to the observed GWB, linearly superimposed on the time-independent, isotropic component described above.
We write this kinematic dipole as the sum of two contributions: one due to the motion of the solar system barycentre (which does not vary appreciably over observational timescales), and the other due to the motion of the Earth around this barycentre (which undergoes annual modulation).
As in~\cite{Planck:2018nkj}, we call these the ``solar dipole'' and ``orbital dipole'', respectively.
Hence the dipole density parameter can be written as\footnote{%
    Note that $\Omega^{(\mathrm{d})}$ is a distribution on the sky, and therefore has units of GWB energy density \emph{per unit solid angle}.
    This amounts to a factor $4\uppi$ difference in normalisation compared to the monopole.}
    \begin{equation}
        \Omega^{(\mathrm{d})}(f,t,\vu*r)=\cald(f)\qty[\vu*r\vdot\vu*r_\sol+\frac{\vu*r\vdot\vb*v_\orb(t)}{v_\sol}],
    \end{equation}
    where $\vu*r_\sol$ is a unit vector pointing in the direction of the solar dipole, $v_\sol$ is the peculiar velocity of the solar system barycentre in that direction, and $\vb*v_\orb$ is the Earth's orbital velocity vector around the barycentre as a function of time.

One can show that, for a power-law monopole signal of the kind we are considering here, the amplitude of the dipole follows the same power law in frequency~\cite{Jenkins:2018nty,Jenkins:2018uac},
    \begin{equation}
        \cald(f)=\cald_\alpha(f/f_\mathrm{ref})^\alpha.
    \end{equation}
In the case of CMB temperature anisotropies, the magnitude of the dipole is trivially related to the monopole by a factor of $v_\sol$, due to the behaviour of the thermal black-body spectrum under Lorentz boosts.
For the GWB this is generally not the case.
In fact, one can show that~\cite{Jenkins:2022chv}
    \begin{equation}
    \label{eq:kd-amplitude}
        \cald(f)=\frac{v_\sol}{4\uppi}\qty(4-\pdv{(\ln f_\mathrm{s})})\Omega(f)+\order{v_\sol^3},
    \end{equation}
    where the derivative should be understood as being taken inside the double integral which defines the monopole $\Omega(f)$ in Eq.~\eqref{eq:monopole}.\footnote{%
    The fact that the derivative is evaluated in this way means that this term is not necessarily proportional to $\alpha$.
    For example, in the case of cosmic strings, the energy radiated by each string loop is \emph{not} scale invariant, so this derivative term is non-zero~\cite{Jenkins:2018nty}.
    The monopole is only (approximately) scale-invariant due to the interplay of the energy spectrum of each loop, the scaling distribution of loop sizes, and the expansion of the Universe.}
(We have included only the linear-order term here, since $v_\sol\sim10^{-3}$ in relativistic units.
See~\cite{Cusin:2022cbb} for a recent discussion of higher-order kinematic effects in the GWB.)

As a simple, specific example, consider the GWB generated by inspiralling compact binaries.
Here, all of the binaries emit GWs with the same $\alpha=2/3$ spectrum, so we find
    \begin{equation}
    \label{eq:cbc-dipole}
        \cald_{2/3}=(5/6\uppi)v_\sol\Omega_{2/3}.
    \end{equation}
This relationship between the dipole and monopole amplitudes is a robust and non-trivial prediction for the compact binary GWB.
A measurement of the dipole amplitude is therefore a valuable consistency check against an observed monopole signal: for example, if the inferred value of $\cald_{2/3}$ were to disagree with Eq.~\eqref{eq:cbc-dipole}, this would point to either a deviation from the $\sim f^{2/3}$ power law expected from inspiralling compact binaries,\footnote{Such deviations could be caused by, e.g., a larger-than-expected contribution from massive binaries which merge within the LIGO/Virgo frequency band~\cite{Callister:2016ewt}, or perhaps additional energy loss channels during the inspiral, such as friction effects from the binary's astrophysical environment~\cite{Cardoso:2019rou} or coupling to fundamental fields beyond the Standard Model~\cite{Huang:2018pbu}.} or a disagreement between the true solar dipole velocity $v_\sol$ and the value inferred by CMB observations, either of which would be a significant and interesting result.

More generally, the fact that each GW source has a different multiplicative factor relating the monopole and dipole amplitudes means that, in principle, a joint measurement of both could reveal the physics of the sources that contribute to the GWB.
This implies that a targeted search for the dipole is extremely useful: despite being an ``extrinsic'' effect due to the observer, it contains valuable astrophysical and cosmological information.

\subsection{The solar dipole}

We now calculate the dipole induced by the motion of the Sun relative to the cosmic rest frame.
In doing so, it is convenient to work in spherical harmonics,
    \begin{equation}
        \Omega^{(\mathrm{d})}_{\ell m}=(f/f_\mathrm{ref})^{-\alpha}\int_{S^2}\dd[2]{\vu*r}Y^*_{\ell m}(\vu*r)\Omega^{(\mathrm{d})}(f,t,\vu*r),
    \end{equation}
    where we have factored out the frequency dependence.

The right ascension and declination of the solar dipole in equatorial coordinates (chosen to match the spherical harmonic decomposition in Eq.~\eqref{eq:ev-correlator}), as inferred by \emph{Planck}~\cite{Planck:2018nkj}, are
    \begin{equation}
        (\alpha_\sol,\delta_\sol)=(167.942\pm0.007,-6.944\pm0.007)\,\mathrm{deg}.
    \end{equation}
Neglecting the orbital dipole for now, this allows us to derive the spherical harmonics associated with the solar dipole,
    \begin{align}
    \begin{split}
    \label{eq:shcs-solar-dipole}
        \Omega^{(\sol)}_{10}&=\sqrt{\frac{4\uppi}{3}}\cos(\uppi/2-\delta_\sol)\cald_\alpha\\
        &=-0.24744\,\cald_\alpha,\\
        \Omega^{(\sol)}_{11}&=-\sqrt{\frac{2\uppi}{3}}\sin(\uppi/2-\delta_\sol)\rme^{-\rmi\alpha_\sol}\cald_\alpha\\
        &=\qty(1.40489+0.30011\,\rmi)\,\cald_\alpha,
    \end{split}
    \end{align}
    with $\Omega^{(\sol)}_{1,-1}=-\Omega^{(\sol)*}_{11}$, and all other harmonics equal to zero.

\subsection{The orbital dipole}

Let us now include the orbital dipole induced by the motion of the Earth relative to the solar system barycentre.
We write the Earth's orbital velocity as
    \begin{equation}
        \vb*v_\orb=\frac{\omega a}{\sqrt{1-e^2}}[-\sin\psi\,\vu*x+(\cos\psi+e)\vu*y],
    \end{equation}
    where $\omega\equiv2\uppi/T_\orb$ is the mean  angular velocity, $T_\orb$ is one sidereal year, $a$ is the Earth's semi-major axis, and $e$ its eccentricity.
The right-handed orthonormal vectors $\vu*x,\vu*y$ are defined so that $\vu*x$ points from the barycentre to Earth's perihelion, and the true anomaly $\psi\qty(t)$ is measured from that point.
The true anomaly does not have a closed-form expression as a function of time for general eccentricity $e>0$.
However, since the Earth's eccentricity is small, $e\sim10^{-2}$, we can perform a series expansion around the circular case~\cite{Murray:2000ssd},
    \begin{align}
    \begin{split}
        \cos\psi&=\cos\omega t+e(\cos2\omega t-1)+\order*{e^2},\\
        \sin\psi&=\sin\omega t+e\sin2\omega t+\order*{e^2},
    \end{split}
    \end{align}
    with the time $t$ chosen such that the Earth is at perihelion at $t=0$.
This gives
    \begin{align}
    \begin{split}
        \vb*v_\orb=&-\omega a(\sin\omega t+e\sin2\omega t)\vu*x\\
        &+\omega a(\cos\omega t+e\cos2\omega t)\vu*y+\order*{e^2}.
    \end{split}
    \end{align}
The spherical harmonics of the orbital dipole therefore read
    \begin{align}
    \begin{split}
    \label{eq:shcs-orbital-dipole}
        \Omega^{(\orb)}_{10}=\sqrt{\frac{4\uppi}{3}}\frac{\omega a}{v_\sol}\cald_\alpha[&-\cos\theta_{\vu*x}(\sin\omega t+e\sin2\omega t)\\
        &+\cos\theta_{\vu*y}(\cos\omega t+e\cos2\omega t)],\\
        \Omega^{(\orb)}_{11}=\sqrt{\frac{2\uppi}{3}}\frac{\omega a}{v_\sol}\cald_\alpha[&+\sin\theta_{\vu*x}\rme^{-\rmi\alpha_{\vu*x}}(\sin\omega t+e\sin2\omega t)\\
        &-\sin\theta_{\vu*y}\rme^{-\rmi\alpha_{\vu*y}}(\cos\omega t+e\cos2\omega t)],
    \end{split}
    \end{align}
    where $\theta\equiv\qty(\uppi/2)-\delta$, and $\alpha_{\vu*x},\delta_{\vu*x}$ and $\alpha_{\vu*y},\delta_{\vu*y}$ are the equatorial coordinates of the unit vectors $\vu*x,\vu*y$, which we compute as~\cite{Murray:2000ssd,Astropy:2013muo,Astropy:2018wqo}
  \begin{align}
    \begin{split}
        \alpha_{\vu*x}&=104.06721\,\mathrm{deg},\ \
        \delta_{\vu*x}=\phantom{1}22.80924\,\mathrm{deg},\\
        \alpha_{\vu*y}&=191.91063\,\mathrm{deg},\ \
        \delta_{\vu*y}=\;-5.11317\,\mathrm{deg}.
    \end{split}
    \end{align}
For the Earth's orbital parameters we take~\cite{Murray:2000ssd}
    \begin{align}
    \begin{split}
        \omega&=1.99102\times10^{-7}\,\mathrm{Hz},\\
        a&=1.49598\times10^{8}\,\mathrm{km},\\
        e&=0.01671,
    \end{split}
    \end{align}
    and for the solar dipole velocity we use the value inferred by \emph{Planck}~\cite{Planck:2018nkj},
    \begin{equation}
        v_\sol=369.82\pm0.11\,\mathrm{km}\,\mathrm{s}^{-1}=(1.2336\pm0.0004)\times10^{-3}.
    \end{equation}
Note that the fractional changes in the overall dipole due to the orbital dipole are of the order $\omega a/v_\sol\approx8\,\%$.

In summary, the spherical harmonics associated with the kinematic dipole (solar plus orbital) are
    \begin{align}
    \begin{split}
    \label{eq:KD_signal}
        \Omega^{(\mathrm{d})}_{10}&=-\cald_\alpha(0.24745+0.06390\sin\omega t+0.00107\sin2\omega t\\
        &\qquad\qquad+0.01469\cos\omega t+0.00025\cos2\omega t),\\
        \Omega^{(\mathrm{d})}_{11}&=\cald_\alpha[1.40489+0.30011\,\rmi\\
        &\qquad\quad-\qty(0.02612+0.10422\,\rmi)\sin\omega t\\
        &\qquad\quad-\qty(0.00044+0.00174\,\rmi)\sin2\omega t\\
        &\qquad\quad+\qty(0.11359-0.02396\,\rmi)\cos\omega t\\
        &\qquad\quad+\qty(0.00190-0.00040\,\rmi)\cos2\omega t].
    \end{split}
   \end{align}
For a fixed power-law index $\alpha$ there is only one free parameter, the amplitude $\cald_\alpha$.

\subsection{Relationship to the angular power spectrum}

The kinematic dipole generates a contribution to the $\ell=1$ multipole of the GWB angular power spectrum,
    \begin{equation}
    \label{eq:C_ell}
        C_\ell=\frac{1}{2\ell+1}\sum_{m=-\ell}^{+\ell}|\Omega_{\ell m}|^2.
    \end{equation}
Using Eqs.~\eqref{eq:shcs-solar-dipole} and~\eqref{eq:shcs-orbital-dipole}, we calculate the size of this contribution, allowing us to compare the results of our search with the upper limits on $C_1$ obtained by the LVK directional GWB search~\cite{KAGRA:2021mth}.

For the solar dipole alone, the corresponding angular power is
    \begin{equation}
    \label{eq:C_1_and_D}
        C_1=\frac{4\uppi}{9}\cald_\alpha^2.
    \end{equation}
Including the orbital dipole gives rise to time-dependent contributions such that $C_1$ is no longer constant.
Many of these time-dependent terms vanish if one integrates over one sidereal year $T_\orb$, leaving just a small enhancement to the $\ell=1$ multipole,
    \begin{equation}
        C_1\approx\frac{4\uppi}{9}\cald_\alpha^2\qty[1+\qty(\frac{\omega a}{v_\sol})^2],
        \label{e:C1_with_correction}
    \end{equation}
    where $(\omega a/v_\sol)^2\approx6.4\times10^{-3}$.
(As above, we have neglected terms of order $e^2$.)

\section{Search method}
\label{sec:Method}

\subsection{Statistics of cross-correlated strain data}

We search for the GWB by cross-correlating data measured by different detectors~\cite{Allen:1997ad,Romano:2016dpx}, as this allows us to differentiate the GW signal from the noise present in each detector.
To do so, we partition each data time series into segments of length $\tau$ (chosen such that $1/\tau$ is much smaller than the GW frequencies we are sensitive to), labelling each segment by its mid-point time $t$.
We then analyse these data segments in the Fourier domain, with $f$ labelling the different frequency bins.
(We can include only positive frequencies without loss of generality.)
For each pair of detectors $(I,J)$, we form the cross-correlation power spectrum estimator
    \begin{equation}
    \label{eq:correlator}
        \hatp_{IJ}(f,t)\equiv\frac{2}{\tau}s_I(f,t)s_J^*(f,t),
    \end{equation}
    where $s_I(f,t)$ are the Fourier-domain strain data from detector $I$.
Assuming zero noise correlation between different detectors~\cite{Meyers:2020qrb}, the expectation value of $\hatp_{IJ}(f,t)$ reads
    \begin{align}
    \begin{split}
    \label{eq:ev-correlator}
        \calp_{IJ}(f,t)&\equiv\ev*{\hatp_{IJ}(f,t)}\\
        &=\delta_{IJ}\caln_I(f,t)+\mathcal{H}_{\alpha}(f)
    \sum_{\ell=0}^\infty\sum_{m=-\ell}^{+\ell}
        \gamma^{IJ}_{\ell m}(f,t)\Omega_{\ell m}(t),
    \end{split}
    \end{align}
    where $\caln_I$ is the power spectral density (PSD) of the noise in detector $I$ (which we allow to be non-stationary), and $\Omega_{\ell m}$ are the spherical harmonics of the GWB.
The $\gamma^{IJ}_{\ell m}$ are the spherical harmonic components of the \emph{overlap reduction function} (ORF) for detector pair $(I,J)$~\cite{Allen:1997ad,Romano:2016dpx}, which describes the average cross-power coherence between the two detectors, as determined by their separation and relative orientations.
Note that both sets of spherical harmonics are time-dependent; for the ORF, this time-dependence is due to the Earth's daily rotation, while for the GWB it is due to the annual modulation of the dipole.
The function
    \begin{equation}
        \mathcal{H}_{\alpha}(f)\equiv\frac{3H_0^2}{2\uppi^2f_\mathrm{ref}^3}(f/f_\mathrm{ref})^{\alpha-3}
    \end{equation}
    translates between the signal PSD and GWB energy density spectrum at frequency $f_\mathrm{ref}$.

Since we are targeting the kinematic dipole, we assume that the spherical harmonics $\Omega_{\ell m}$ in Eq.~\eqref{eq:ev-correlator} are given by those we calculated in section~\ref{sec:KD-model}.
In what follows, it is convenient to define the signal template
    \begin{equation}
    \label{eq:template}
        \calt_{\alpha,IJ}\qty(f,t)\equiv\frac{\mathcal{H}_{\alpha}(f)}{\cald_\alpha}\sum_{\ell m}\gamma^{IJ}_{\ell m}(f,t)\Omega_{\ell m}^{(\mathrm{d})}(t),
    \end{equation}
    which encodes the time- and frequency-dependence of the response of the detector pair to a dipole with spectral index $\alpha$.
This allows us to write Eq.~\eqref{eq:ev-correlator} as
    \begin{equation}
        \calp_{IJ}=\delta_{IJ}\caln_I+\cald_\alpha \calt_{\alpha,IJ}.
    \end{equation}
(Note that there is no implied sum over $\alpha$.)

Assuming the data $s_I(f,t)$ are Gaussian, we calculate the covariance of different correlators~\eqref{eq:correlator} using Isserlis' theorem, as
    \begin{align}
   \begin{split}
    \label{eq:cov-correlator}
        \cov&[\hatp_{IJ}(f,t),\hatp_{I'J'}(f',t')]
        =\delta_{ff'}\delta_{tt'}\calp_{II'}(f,t)\calp^*_{JJ'}(f,t)\\
        &\qquad\qquad\qquad=\delta_{ff'}\delta_{tt'}\delta_{II'}\delta_{JJ'}\caln_I\caln_J+\order{\cald_\alpha}.
    \end{split}
    \end{align}

\subsection{Maximum-likelihood dipole estimator}
The statistics of our complex strain data are defined by a circularly-symmetric normal log-likelihood function,
    \begin{equation}
        \call\qty(s|\cald_\alpha)=-\sum_{f,t}\qty[\ln\qty(\det\qty(\uppi\mathsf{C}))+\sum_{I,J}s_I^*\mathsf{C}^{-1}_{IJ}s_J],
    \end{equation}
    where the sum is over data segments and positive frequency bins.
Here we have defined the covariance matrix,
    \begin{equation}
    \label{eq:cov}
        \mathsf{C}_{IJ}=\ev{s_Is_J^*}=\frac{\tau}{2}(\delta_{IJ}\caln_I+\cald_\alpha \calt_{\alpha,IJ}),
    \end{equation}
    whose inverse can be written as
    \begin{equation}
    \label{eq:inv-cov}
        \mathsf{C}^{-1}_{IJ}=\frac{2}{\tau}\qty(\frac{\delta_{IJ}}{\caln_I}-\frac{\cald_\alpha\calt_{\alpha,IJ}}{\caln_I\caln_J})+\order{\cald_\alpha^2}.
    \end{equation}

We construct an estimator for the dipole amplitude $\cald_\alpha$ by finding the particular value $\hatd_\alpha(s)$ that maximises this likelihood function for a given realisation of the strain data.
    The simplest way to do this is to find the value for which the derivative $\pdv*{\call}{\cald_\alpha}$ vanishes.
Using the standard identities
    \begin{align}
    \begin{split}
        &\pdv{x}\ln(\det\mathsf{M})=\Tr{\mathsf{M}^{-1}\pdv{\mathsf{M}}{x}},\\
        &\pdv{x}\mathsf{M}^{-1}=-\mathsf{M}^{-1}\pdv{\mathsf{M}}{x}\mathsf{M}^{-1},
    \end{split}
    \end{align}
    we therefore write
    \begin{align}
    \begin{split}
    \label{eq:likelihood-derivative}
        \pdv{\call}{\cald_\alpha}&=\sum_{f,t}\bigg[-\sum_{I,J}\mathsf{C}^{-1}_{IJ}\pdv{\mathsf{C}_{JI}}{\cald_\alpha}\\
        &\qquad\qquad+\sum_{I,J,K,L}\mathsf{C}^{-1}_{IJ}{\pdv{\mathsf{C}_{JK}}{\cald_\alpha}}\mathsf{C}^{-1}_{KL}\frac{\tau}{2}\hatp_{LI}\bigg],
    \end{split}
    \end{align}
    where we have replaced the strain data $s_I$ with the strain power spectrum estimator $\hatp_{IJ}=(2/\tau)s_Is^*_J$.

Note that Eq.~\eqref{eq:likelihood-derivative} includes the auto-power estimators $\hatp_{II}$.
In principle, these include a contribution from the dipole signal, as well as from the noise auto-power in each detector (cf. Eq.~\eqref{eq:cov}).
In practice, this signal contribution is much smaller than the uncertainty on the noise power, so the auto-correlation terms are not useful in constructing our estimator.
We therefore `average them out', replacing them with their mean values, and keeping only the cross-correlation terms.
Inserting Eqs.~\eqref{eq:cov} and~\eqref{eq:inv-cov} into Eq.~\eqref{eq:likelihood-derivative}, we therefore obtain
    \begin{equation}
    \label{eq:likelihood-derivative-simplified}
        \pdv{\call}{\cald_\alpha}=\sum_{f,t}\sum_{I\ne J}\frac{\calt_{\alpha,IJ}^*(\hatp_{IJ}-\cald_\alpha \calt_{\alpha,IJ})}{\caln_I\caln_J}+\order{\cald_\alpha^2}.
    \end{equation}

This can be written more compactly by defining an inner product,
    \begin{equation}
    \label{eq:inner-prod}
        (\mathcal{A}|\mathcal{B})\equiv\sum_{f,t}\sum_{I\ne J}\frac{\mathcal{A}^*_{IJ}\mathcal{B}_{IJ}}{\caln_I\caln_J}=2\sum_{f,t}\sum_{I>J}\Re{\frac{\mathcal{A}^*_{IJ}\mathcal{B}_{IJ}}{\caln_I\caln_J}},
    \end{equation}
    which describes the noise-weighted coherence between two stochastic cross-power signal models, $\mathcal{A}_{IJ}(f,t)$ and $\mathcal{B}_{IJ}(f,t)$.
(The second equality here comes from the fact that exchanging the two detector indices $I,J$ is equivalent to taking a complex conjugate.)
We write the associated norm as
    \begin{equation}
        \norm{\mathcal{A}}^2\equiv(\mathcal{A}|\mathcal{A}).
    \end{equation}
Equation~\eqref{eq:likelihood-derivative-simplified} then becomes
    \begin{equation}
        \pdv{\call}{\cald_\alpha}=(\calt_\alpha|\hatp-\cald_\alpha \calt_\alpha)+\order{\cald_\alpha^2}.
    \end{equation}
Setting this to zero, and keeping only the leading-order term in $\cald_\alpha$, we therefore obtain our maximum-likelihood estimator for the dipole amplitude,
    \begin{equation}\label{eq:pt_Estimate}
        \hatd_\alpha(s)\equiv\frac{(\calt_\alpha|\hatp(s))}{\norm{\calt_\alpha}^2}\,,
    \end{equation}
where we have emphasised the data-dependence of the power spectrum estimator $\hatp_{IJ}\equiv(2/\tau)s_Is_J^*$ on the right-hand side of the above equation.

\subsection{Sampling distribution of the estimator}

A simple calculation shows that our dipole estimator is unbiased,
    \begin{equation}
        \ev*{\hatd_\alpha}=\frac{\ev*{(\calt_\alpha|\hatp)}}{\norm{\calt_\alpha}^2}=\frac{(\calt_\alpha|\cald_\alpha \calt_\alpha)}{\norm{\calt_\alpha}^2}=\cald_\alpha.
    \end{equation}
To calculate the variance of the estimator, note that
    \begin{align}
    \begin{split}
        \var[(\calt_\alpha|\hatp)]&=\sum_{f,f,t',t'}\sum_{I\ne J,I'\ne J'}\frac{\calt^*_{\alpha,IJ}\calt_{\alpha,I'J'}\cov[\hatp_{IJ},\hatp_{I'J'}]}{\caln_I\caln_J\caln_{I'}\caln_{J'}}\\
        &=\sum_{f,t}\sum_{I\ne J}\frac{|\calt_{\alpha,IJ}|^2}{\caln_I\caln_J}(1+\order{\cald_\alpha})\\
        &=\norm{\calt_\alpha}^2(1+\order{\cald_\alpha}),
    \end{split}
    \end{align}
    where we have used Eq.~\eqref{eq:cov-correlator} in the second equality.
We therefore have
    \begin{equation}
    \label{eq:dipole-variance}
        \var[\hatd_\alpha]=\frac{\var[(\calt_\alpha|\hatp)]}{\norm{\calt_\alpha}^4}=\frac{1}{\norm{\calt_\alpha}^2}(1+\order{\cald_\alpha}).
    \end{equation}
(In what follows we will no longer keep track of $\order{\cald_\alpha}$ corrections, as the leading-order term here is sufficient for our analysis.)

Note that, in the weak-signal limit, Eq.~\eqref{eq:dipole-variance} saturates the Cram\'er-Rao bound, which is given here by
    \begin{equation}
        \var[\hat\cald_\alpha]\ge\frac{1}{\ev*{(\pdv*{\call}{\cald_\alpha})^2}}=\frac{1}{\var[(\calt_\alpha|\hatp)]}=\frac{1}{\norm{\calt_\alpha}^2}.
    \end{equation}
This shows that $\hatd_\alpha$ is the minimum-variance unbiased estimator for $\cald_\alpha$ (as expected for a maximum-likelihood estimator).

We have assumed that the strain data $s$ are Gaussian random variables, which means that the strain power spectrum estimators $\hatp$ are definitively \emph{non}-Gaussian (rather, they follow $\chi^2$ distribution).
However, our dipole estimator is a sum over
    \begin{equation}
        N\equiv\sum_{f,t}\sum_{I\ne J}
    \end{equation}
    independent real degrees of freedom.
For large $N$, this means that $\hatd_\alpha$ is asymptotically Gaussian by the central limit theorem.
For the LIGO/Virgo O3 dataset we have $N\sim10^{10}$, so the Gaussian approximation is extremely accurate.\footnote{
    This assumes $\sim1\,\mathrm{yr}$ of observation with three detectors, using $\tau=192\,\mathrm{s}$ intervals, with frequency resolution $1/32$~Hz in the band 20--1726~Hz~\cite{KAGRA:2021kbb}.}
This means that the mean and variance fully specify the sampling distribution of $\hatd_\alpha$.

\subsection{Signal-to-noise ratio and upper limits}

It is convenient to define the signal-to-noise ratio (SNR) of our search,\footnote{%
    This is completely analogous to the standard matched-filter SNR one usually defines for coherent GW signals---the difference here being that the inner product captures \emph{strain power coherence} over a detector network, rather than \emph{strain coherence} in a single detector.}
    \begin{equation}
    \label{eq:snr}
        \rho_\alpha\equiv\frac{\hatd_\alpha}{\sqrt{\var[\hatd_\alpha]}}=\frac{(\calt_\alpha|\hatp)}{\norm{\calt_\alpha}}.
    \end{equation}
Given the results above, we see that in the absence of a signal ($\cald_\alpha=0$) this follows a standard normal distribution, $\rho_\alpha\sim\mathrm{N}(0,1)$, which is very convenient for interpreting the statistical significance of a given search result.
If a signal is present, then the expected SNR, $\ev{\rho_\alpha}=\cald_\alpha\norm{T_\alpha}$, scales as the square root of the total observation time, due to the increasing number of terms that are summed over in the inner product.
We therefore recover the standard $\mathrm{SNR}\propto\sqrt{T_\obs}$ scaling from other stochastic searches.

If we observe a SNR, $\rho_{\alpha,\obs}$, that is not high enough to claim a detection, then we can instead set an upper limit by noting that, for a given value of $\cald_\alpha$,
    \begin{equation}
    \label{eq:Pr-l>l*}
        \mathrm{Pr}(\rho_\alpha\ge\rho_{\alpha,\obs}|\cald_\alpha)=\frac{1}{2}\qty[1+\erf\qty(\frac{\cald_\alpha\norm{\calt_\alpha}-\rho_{\alpha,\obs}}{\sqrt{2}})].
    \end{equation}
For a given confidence level $p\in\qty(0,1)$, the corresponding upper limit on $\cald_\alpha$ is obtained by setting \eqref{eq:Pr-l>l*} equal to $p$ and then solving for $\cald_\alpha$, e.g., for $p=0.95$,
    \begin{equation}
    \label{eq:UL}
        \cald_\alpha^\mathrm{95UL}\equiv\frac{\rho_{\alpha,\obs}+\sqrt{2}\erf^{-1}\qty(2\times 0.95-1)}{\norm{\calt_\alpha}}.
    \end{equation}

\subsection{Geometric interpretation}

The inner product that we have introduced in Eq.~\eqref{eq:inner-prod} defines a $N$-dimensional vector space, in which individual vectors represent different possible cross-power signal models for our detector network.
In light of this, our search SNR~\eqref{eq:snr} has the simple geometric interpretation of projecting the cross-correlated data onto the ``$\calt_\alpha$ direction'' in this space; i.e., the direction corresponding to our kinematic dipole signal template.

Using Eqs.~\eqref{eq:ev-correlator} and~\eqref{eq:cov-correlator} we see that, in the absence of a signal, the cross-correlated data vector $\hatp$ has a mean square length of
    \begin{equation}
        \ev*{\norm*{\hatp}{}^2}=N,
    \end{equation}
 as one would expect for a random walk of $N$ steps of unit length.
This is because each term which is summed over in the inner product has unit mean,
    \begin{equation}
        \frac{\ev*{|\hatp_{IJ}|^2}}{\caln_I\caln_J}=1,
    \end{equation}
    i.e., for each of the $N$ dimensions of the signal space the data vector takes a statistically independent random step with zero mean and unit variance.
The detection statistic measures the projection of this random walk onto one of these $N$ directions (that parallel to the template vector $\calt_\alpha$), which explains why $\rho_\alpha\sim\mathrm{N}\qty(0,1)$ in the absence of a signal.
If a signal is present, then it causes the random walk to travel preferentially in the $\calt_\alpha$ direction and gives a positive offset in the corresponding inner product.

\subsection{Bias from ignoring the orbital dipole}
\label{sec:bias}

\begin{figure}[t!]
    \centering
    \subfigure{
        \includegraphics[width=0.45\textwidth]{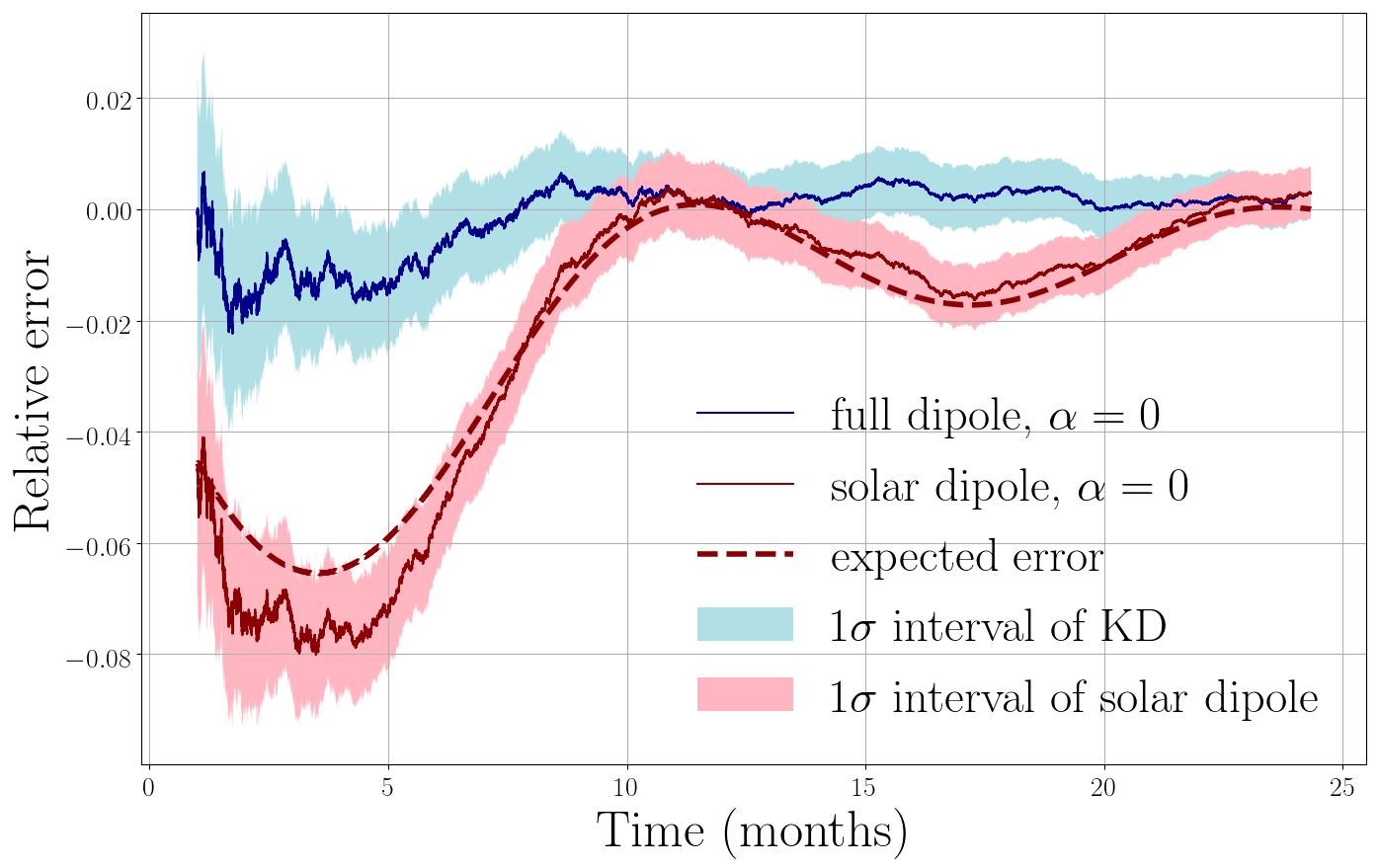}
        \label{fig:subfig1}
    }
    \subfigure{
        \includegraphics[width=0.45\textwidth]{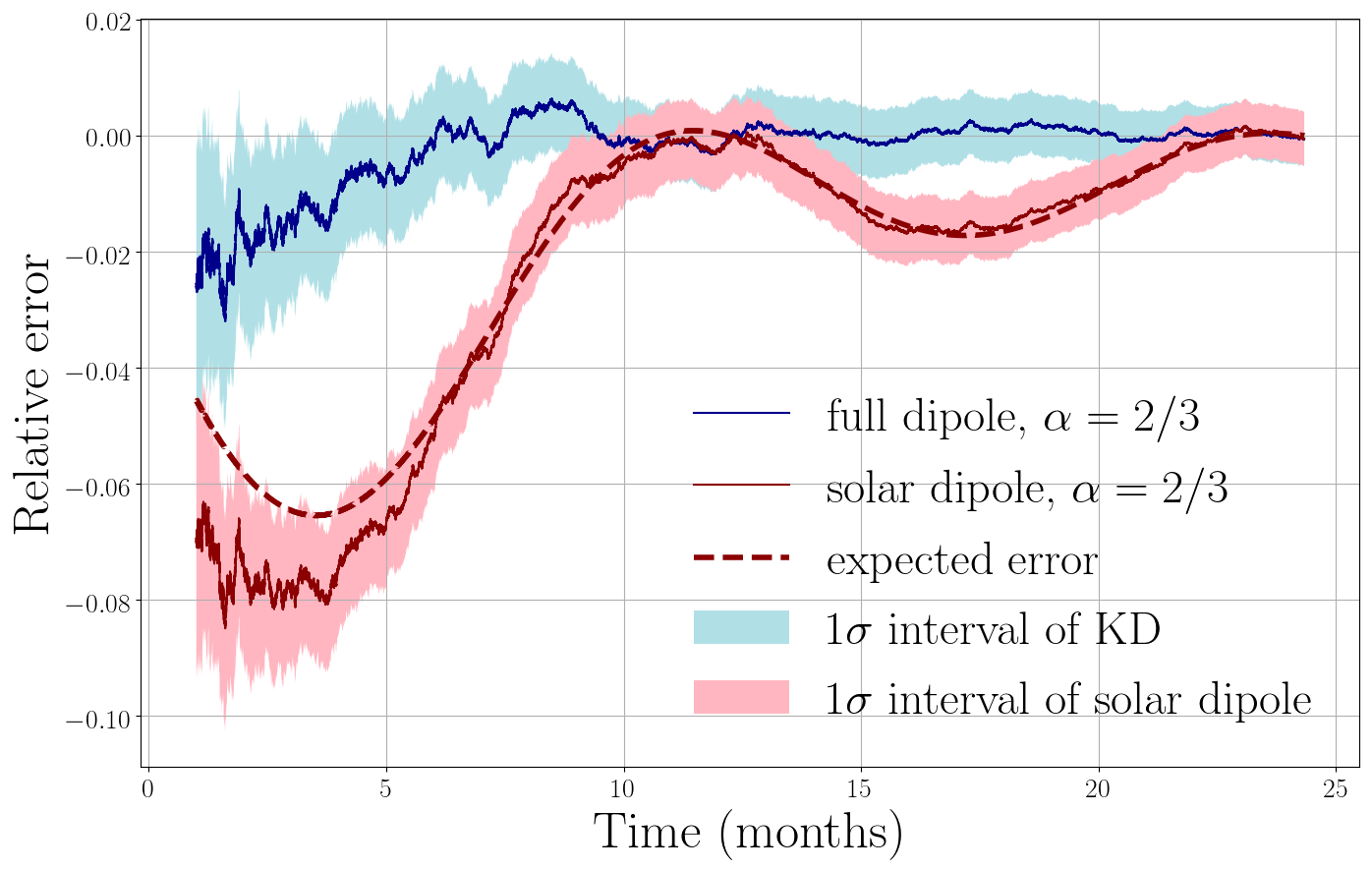}
        \label{fig:subfig2}
    }
    \subfigure{
        \includegraphics[width=0.45\textwidth]{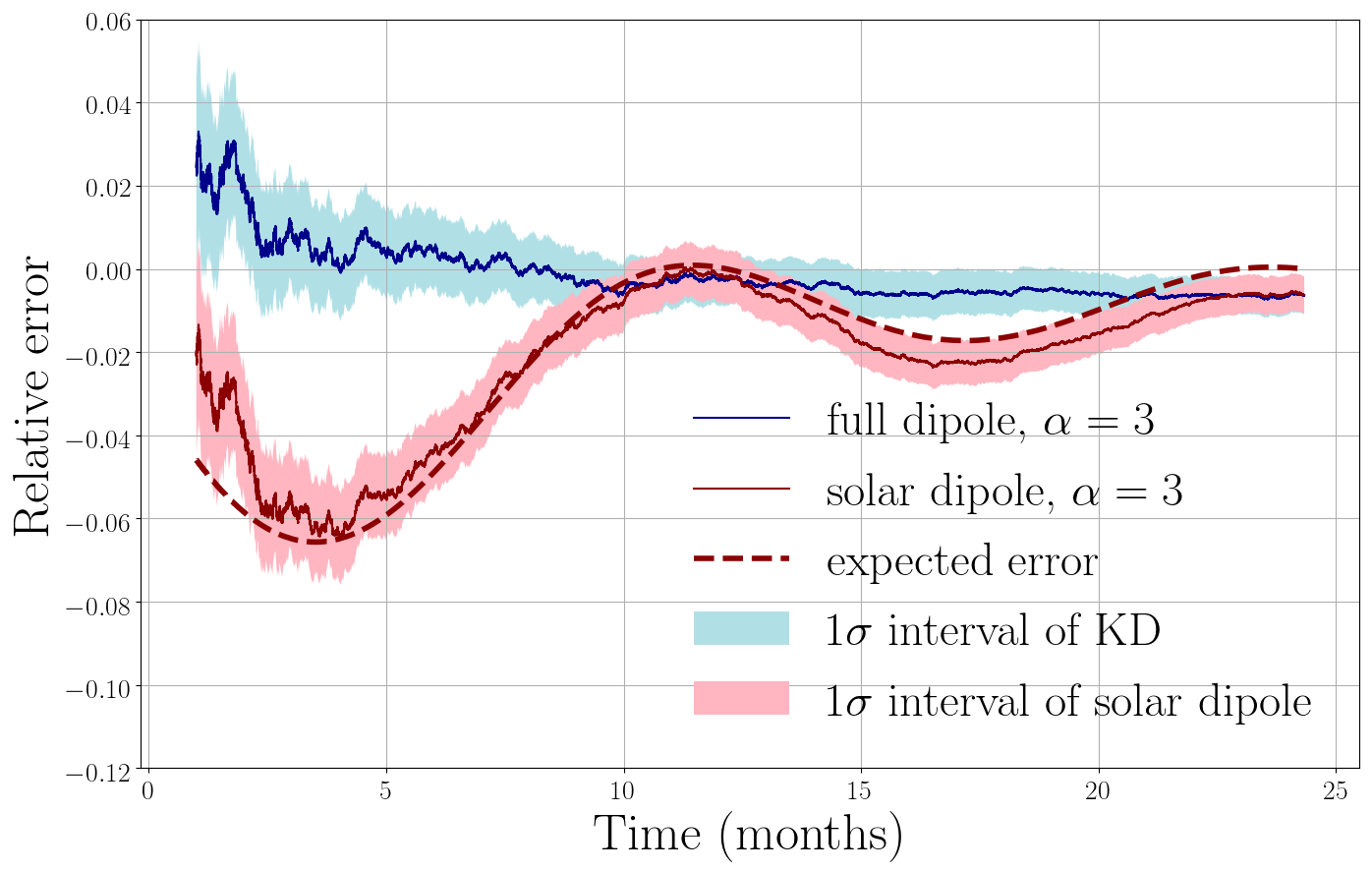}
        \label{fig:subfig3}
    }
    \caption{
    Relative error of our targeted kinematic dipole search as a function of observing time for simulated data containing a kinematic dipole signal $\cald_\alpha = 3.1 \times 10^{-8} $ for $\alpha = 0 $ (top panel), $\cald_\alpha=2.3 \times 10^{-8} $ for $\alpha = 2/3 $ (middle panel), and $\cald_\alpha= 4 \times 10^{-9} $ for $\alpha = 3 $ (bottom panel).
    The blue and red traces correspond to analyses using the full (time-dependent) dipole template and the (time-independent) solar dipole template, respectively.
    While the relative errors for the full dipole template analyses converge to zero rapidly, those obtained using the solar-dipole template show a relative error consistent with the expected error \eqref{eq:relative_error} (dashed red line).
    }
    \label{fig:new_injections}
\end{figure}

One advantage of the method described above over existing searches is that it allows for a time-dependent signal model, enabling us to capture the annual modulation of the dipole signal due to the Earth's orbital motion.
We can calculate the bias which would be introduced if one ignored this orbital contribution by writing
    \begin{equation}\label{eq:relative_error}
        \text{relative error}=\frac{\hatd^{(\sol)}_\alpha-\cald_\alpha}{\cald_\alpha}=\frac{(\calt_\alpha^{(\sol)}|\hatp)}{\cald_\alpha\norm*{\calt_\alpha^{(\sol)}}{}^2} - 1,
    \end{equation}
    where $\cald_\alpha$ is the true value of the dipole amplitude, and $\hatd_\alpha^{(\sol)}$ is the value inferred using a template, $\calt_\alpha^{(\sol)}$, which only includes the solar dipole.
Since $\ev*{\hatp}=\cald_\alpha \calt_\alpha$, we find that
    \begin{equation}
        \ev{\text{relative error}}=\frac{(\calt_\alpha^{(\sol)}|\calt_\alpha)}{\norm*{\calt_\alpha^{(\sol)}}{}^2}-1.
    \end{equation}
In section~\ref{sec:results_simulated} and figure~\ref{fig:new_injections} we will further investigate this relative error using simulated data.

\begin{table*}[t!]
\centering
\begin{tabular}{lllllll}
\toprule
$\alpha$ ~ &
~ $\rho_{\alpha,\obs}$ ~ & ~
$\hatd_\alpha$  ~ & ~
$\sigma_{\cald_\alpha}$  ~ & ~
$C^{95\rm UL}_1$ ~ & ~
$C^{95\rm UL}_1$ (SHD) \\ \hline
0 ~ & ~  $0.0579$ ~ & ~ $6.58 \times 10^{-11} $ ~ & ~ $1.14 \times 10^{-9} $ ~ & ~ $5.23 \times 10^{-18} $ ~ & ~ $ 1.88 \times 10^{-18} $\\
2/3 ~ & ~ $ 0.442 $ ~ & ~ $3.32 \times 10^{-10}$  ~ & ~ $7.51 \times 10^{-10} $ ~  & ~ $3.43 \times 10^{-18} $ ~ & ~ $ 1.38 \times 10^{-18} $\\
3 ~ & ~ $ 0.426 $ ~ & ~ $ 4.09\times 10^{-11}$  ~ & ~ $9.58 \times 10^{-11} $ ~ & ~ $ 5.50 \times 10^{-20} $ ~ & ~ $ 7.22 \times 10^{-20} $ \\
\toprule
\end{tabular}
\caption{
Results of searching for the kinetic dipole using data from the first three observing runs of the Advanced LIGO and Virgo detectors.
Columns from left to right are:
(i) spectral indices of the GW background energy-density spectrum,
(ii) observed values of the detection statistic $\rho_{\alpha,\obs}$,
(iii) point estimates of the kinematic dipole moment $\hatd_\alpha$,
(iv) measurement uncertainties of the kinematic dipole moment $\sigma_{\cald_\alpha}$,
(v) 95\% confidence-level upper limits on the dipole angular power spectrum component $C_1$, converted from the measurement of $\cald_\alpha$ as described in the main text, and
(vi) 95\% confidence-level upper limits on $C_1$, as determined using the spherical-harmonic decomposition (SHD) approach~\cite{KAGRA:2021mth, O3_SHD}. }
\label{tab:Results}
\end{table*}

\section{Application to real and simulated data}
\label{sec:PE}

\subsection{Data processing}

When applying our data analysis pipeline to either real or (semi-realistic) simulated data, several standard data-processing procedures need to be performed.
First, we divide the time-domain data into shorter segments, typically of duration $192$ seconds.
To suppress spectral leakage, we Hann window the data, overlapping the windowed segments by 50\% to avoid losing roughly half of the data due to the effects of the tapering.
The windowed data are then transformed into the frequency-domain data via the short-time Fourier transform and coarse-grained to a frequency resolution of $1/32$ Hz as described in~\cite{LIGOScientific:2016jlg, LIGOScientific:2019vic, KAGRA:2021kbb, LIGOScientific:2016nwa, LIGOScientific:2019gaw, KAGRA:2021mth}.
Since the data have been windowed, the cross-correlation power estimate~\eqref{eq:pt_Estimate} and its variance~\eqref{eq:dipole-variance} need to be multiplied by windowing factors~\cite{LIGOScientific:2003jxj}.
Second, to avoid a bias when computing the variance, we multiply $\sigma_{\cald_\alpha}$ by a bias factor of 1.0504~\cite{bias_factor}.
The detection statistic and variance of the overlapped data segments are then optimally combined as described in~\cite{Overlapping_windows}.

\subsection{Simulated data}
\label{sec:results_simulated}

We first validate our analysis method and the corresponding data analysis pipeline by performing a targeted dipole search on mock data which contain a GWB with a known kinematic dipole.
For the purpose of this exercise, we simulate a dipole with $\cald_\alpha=3.1\times10^{-8}$, $2.3\times10^{-8}$, and $4\times10^{-9}$ respectively for $\alpha=0$, $2/3$,  and $3$.
These values of $\cald_\alpha$ are chosen so that each dipole has approximately the same accumulated SNR.
The mock data cover two years, with start time equal to the start of LIGO/Virgo's third observing run (O3).
To ensure the gravitational-wave intensity is positive for all sky directions, we include a monopole moment with value $2 \sqrt{4 \pi} \cald_\alpha$.
We estimate the dipole moment for each spectral index using both the full time-dependent dipole template~\eqref{eq:KD_signal} and the time-independent solar-dipole template~\eqref{eq:shcs-solar-dipole}.

Figure~\ref{fig:new_injections} shows the relative error of the point estimate, $(\hat{\cald}_\alpha-\cald_\alpha)/\cald_\alpha$, as a function of observation time, as obtained using the full dipole template (solid blue line) and the solar dipole template (solid red line).
The shaded regions show the $1\sigma$ uncertainty in each case, $1/\norm{\calt_\alpha}$.

From Figure~\ref{fig:new_injections}, we can draw the following two conclusions:
First,  the relative error using the full-dipole template converges to zero very rapidly.
This suggests that our data analysis pipeline can accurately and swiftly recover the injected value of $\cald_\alpha$ for all three spectral indices.
Second, the expected errors predicted using Eq.~\eqref{eq:relative_error} (dashed red lines) are consistent with the measured errors within $\sim\!1\sigma$, which suggests that the measured relative errors are as expected.
Thus, our analysis method can accurately and efficiently measure the dipole moment using the full time-dependent template, avoiding the bias that would result if we ignore the orbital dipole, which can be as large as $\order{10\%}$.


\subsection{LIGO/Virgo data}

Having validated our data analysis method and pipleline, we apply the formalism described in the previous sections to search for evidence of the kinematic dipole using data taken by the Advanced LIGO and Virgo detectors during their first three observing runs (O1, O2, and O3)~\cite{LIGOScientific:2016jlg,LIGOScientific:2019vic,KAGRA:2021kbb,LIGOScientific:2016nwa,LIGOScientific:2019gaw,KAGRA:2021mth}.
To avoid data contamination due to narrow-band noise sources (harmonics of the 60~Hz and 50~Hz power mains, resonances due to mirror suspensions, calibration lines, etc.), the frequencies corresponding to instrumental lines~\cite{O2_directional_notch_list,O3_SHD} are notched.
Data segments which contain non-Gaussian noise artefacts (such as glitches) or resolvable signals from individual GW events are also excluded from the analysis (see also~\cite{LIGOScientific:2016gtq,Thrane:2013npa,Thrane:2014yza,LIGO_data_cut_04}).
For simplicity, we neglect the impact of calibration uncertainty in the LIGO/Virgo interferometers in this analysis. However, for future applications of our pipeline it is straightforward to perform a Bayesian marginalisation over this uncertainty using existing tools~\cite{KAGRA:2021kbb,KAGRA:2021mth}. This marginalisation typically increases the final uncertainty (and associated upper limits) by at most a few percent.

Table~\ref{tab:Results} summarizes the results of our search. The columns show the three spectral indices $\alpha$ of the GW energy-density spectrum searched for, the observed signal-to-noise ratio $\rho_{\alpha,\obs}$, the corresponding point estimates for the dipole amplitude $\hat{\cal D}_{\alpha}$ and their uncertainties $\sigma_{\cal D_{\alpha}}$, and the 95\% confidence upper limits (ULs) for $C_1$ calculated for our kinematic dipole search and (for comparison) the standard LIGO/Virgo spherical harmonic search.
Since $|\rho_{\alpha,\obs}|$ is significantly less than unity, we conclude that we have not detected a kinematic dipole signal, which is consistent with the non-detection of a GWB reported in~\cite{KAGRA:2021kbb, KAGRA:2021mth}.
Our results for $\cald_\alpha$ can be converted to an UL on the dipole-angular power spectrum component $C_1$ by simply using Eq.~\eqref{eq:C_1_and_D},
\begin{equation}
    C_1^\mathrm{95UL}=\frac{4\uppi}{9}(\cald_\alpha^\mathrm{95UL})^2.
\end{equation}
(This is consistent with using Eq.~\eqref{e:C1_with_correction} to less than 1\%.)
These 95\% confidence ULs can be compared to similar ULs calculated using the spherical harmonic decomposition (SHD) method
(right-most column)~\cite{KAGRA:2021mth}.%
\footnote{The values of $C^{95\rm UL}_1$~(SHD) listed in the last column are not the square of $C_{1}^{1/2}$ from Figure 4 of~\cite{KAGRA:2021mth}, which take into account detector calibration error.
Since we do not include the effect of calibration error in our measurement, we also do not include it for the SHD analysis.  Instead, we use
\begin{equation}
    \begin{aligned}
    C_1^{95\rm UL} ({\rm SHD}) &= \hat{C}_1 +\sqrt{2}{\rm erf}^{-1}(2\times 0.95-1)\,\sigma_{C_1}\\
    &= \hat{C}_1 +1.64\,\sigma_{C_1}\,,
\nonumber
\end{aligned}
\end{equation}
where $\hat C_1$ and $\sigma_{C_1}$ are taken from
\cite{O3_SHD}.}
We see that our results are consistent with the existing measurement of anisotropies in the GWB.

\begin{figure}[t!]
    \centering
    \includegraphics[width=0.48\textwidth]{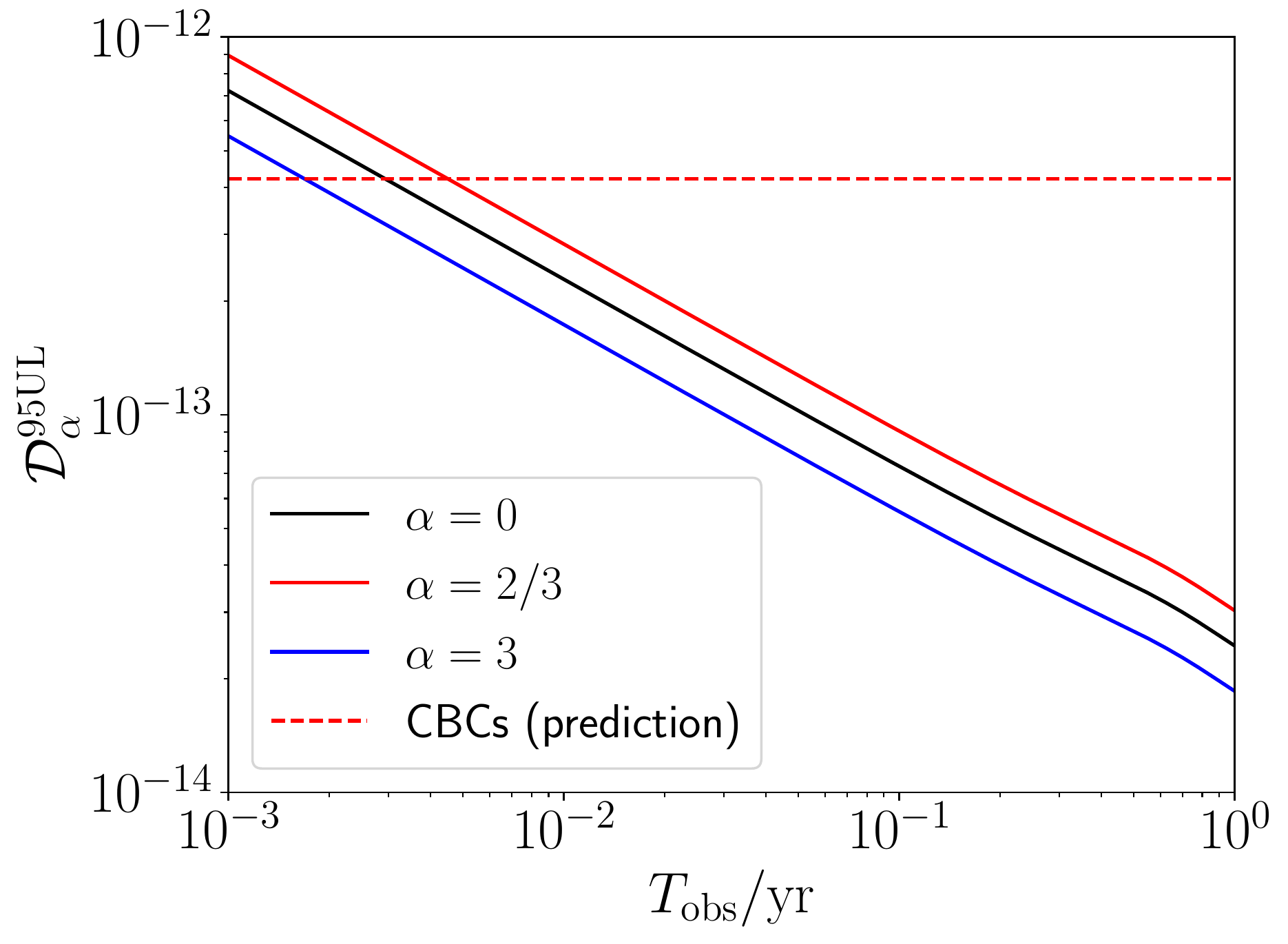}
    \caption{Forecast upper limits (95\% confidence) on the dipole amplitude with a third-generation interferometer network (Einstein Telescope + two Cosmic Explorers).
    The predicted dipole from compact binary coalescences is, in principle, accessible with less than two days of data.}
    \label{fig:forecasts}
\end{figure}

\subsection{Forecasts for third-generation interferometers}

Given the null results reported above, it is interesting to ask how much we can expect our sensitivity to improve with next-generation gravitational-wave observatories such as Einstein Telescope~\cite{Punturo:2010zz} and Cosmic Explorer~\cite{Reitze:2019iox}.
(For Einstein Telescope we assume an L-shaped geometry, and use the sensitivity curve available at~\cite{ETcurve}.)
It is straightforward to forecast the expected upper limits obtained by a given interferometer network using Eq.~\eqref{eq:UL}.
In figure~\ref{fig:forecasts} we compute these forecasts as a function of observing time for a third-generation network consisting of one Einstein Telescope and two Cosmic Explorers.
For the purposes of computing the overlap reduction functions, we assume that Einstein Telescope is located at the current Virgo site, while the two Cosmic Explorers are located at the current sites of the two LIGO interferometers; however, we do not expect our results to be overly sensitive to this choice.
We find that this network will be sensitive to $\cald\approx3\times10^{-14}$ after one year of observations, with some slight variation between different spectral indices---an improvement of three orders of magnitude over our current constraints from LIGO/Virgo.

In figure~\ref{fig:forecasts} we also show the predicted dipole from inspiralling compact binaries~\eqref{eq:cbc-dipole}, whose amplitude we take as $\cald_{2/3}\approx4\times10^{-13}$~\cite{Jenkins:2022chv}.
(This value is based on our current understanding of the likely amplitude of the monopole component, and is therefore subject to some modelling uncertainty, but should be robust to within an order of magnitude.)
We see that our third-generation network is expected to detect this dipole extremely quickly, with less than two days of data.
This shows that future gravitational-wave observatories will be capable of measuring this particular dipole with very high precision, enabling robust tests of our understanding of the source population, and of cosmology more broadly (i.e. by testing whether the dipole direction and amplitude are consistent with that of the CMB, or whether they exhibit anomalies similar to those discussed in~\cite{Secrest:2020has,Dalang:2021ruy,Secrest:2022uvx}).\footnote{Note that we have not investigated here the impact of shot noise on the kinematic dipole from compact binaries~\cite{Jenkins:2019nks,Jenkins:2019uzp}; we leave this for future work.}

\section{Conclusion}
\label{sec:conclusion}

In this paper, we have constructed a targeted search pipeline for the kinematic dipole of the gravitational-wave background induced by the motion of the Earth relative to the cosmic rest frame, accounting for both the motion of the solar system barycentre and the Earth's orbit with respect to this.
The former is the main contribution to the induced dipole, while the latter gives rise to an annual modulation, which we incorporate by generalising the existing search formalism.
By applying our pipeline to mock data, we have shown that we can accurately and efficiently measure the simulated dipole moment, whereas the existing methods, by neglecting the orbital dipole of the dipole signal, result in a bias as large as $\order{10\%}$.
A further advantage of our method over existing searches is that, by constructing an explicit signal template on the sky, we avoid various regularisation issues associated with generic spherical harmonic searches due to effective ``blind spots'' in the response of the detector network.
Instead, our search returns a single, unambiguous measurement of the dipole amplitude.
Our analysis can also be straightforwardly adapted to perform a targeted search for any time- and frequency-dependent stochastic signal with known sky pattern, e.g. GW emission from the Galactic Centre~\cite{Talukder:2014eba,Agarwal:2022lvk}.
It is also straightforward to replace the \emph{Planck} values we have assumed for the dipole magnitude and direction with any other desired values, which may be relevant given the claimed tension between \emph{Planck} and various lower-redshift surveys~\cite{Secrest:2020has,Dalang:2021ruy,Secrest:2022uvx}.

We have applied our pipeline to data from LIGO/Virgo's first three observing runs, and find no evidence for a dipole signal.
However, looking forward to upcoming third-generation interferometers, we have shown that our search will be sensitive to dipoles as small as $\cald_\alpha\sim10^{-14}$, providing a novel probe of cosmology and gravitational-wave source physics.
In particular, for GWB signals with unknown source physics, a joint measurement of the monopole and the dipole will allow us to extract information about the emission spectra of the sources, whereas for GWB signals that are well-understood (e.g. the signal from inspiralling compact binaries), we can instead attempt to provide an independent measurement of our peculiar motion with respect to the cosmic rest frame, allowing us to weigh in on the ``dipole tension'' discussed above.

We have not considered here the impact of shot noise, which will be important for the kinematic dipole from compact binaries~\cite{Jenkins:2019uzp}.
However, in the context of third-generation interferometers it is expected that many of the binaries which make up this ``stochastic'' signal will in fact be individually resolvable, allowing us to coherently subtract them and search for underlying cosmological signals~\cite{Zhu:2012xw,Regimbau:2016ike}, to which our dipole search can be applied.
Alternatively, it may be possible to mitigate the effects of shot noise by optimally combining data from different time segments~\cite{Jenkins:2019nks} and frequency bins~\cite{DallArmi:2022wnq}.
We leave a fuller exploration of these issues for future work.

\begin{acknowledgments}
A.K.W.C.~thanks Andrew Matas, Erik Floden and Patrick Meyers for help in analysing mock and actual data of the Advanced LIGO and Advanced Virgo detectors, and Stuart Anderson for coordinating the file I/O on the CIT clusters.
The authors acknowledge use of data, software and web tools from the GW Open Science Center (https://www.gw-openscience.org), a service of LIGO Laboratory, the LIGO Scientific Collaboration and the Virgo Collaboration.
The authors are grateful for computational resources provided by the LIGO Laboratory and supported by National Science Foundation Grants PHY-0757058 and PHY-0823459.

A.K.W.C.~is supported by the Hong Kong Scholarship for Excellence Scheme (HKSES).
J.D.R. is supported in part by start-up funds provided by Texas Tech University and National Science Foundation grant PHY-2207270.
M.S. is supported in part by the Science and Technology Facility Council (STFC), United Kingdom, under the research grant ST/P000258/1.
This work was partly enabled by the UCL Cosmoparticle Initiative.

This paper has a LIGO document number LIGO-P2200215 and a KCL document number KCL-PH-TH/2022-38.
This material is based upon work supported by NSF's LIGO Laboratory which is a major facility fully funded by the National Science Foundation. This paper has an Einstein Telescope number ET-0169A-22.
\end{acknowledgments}

\bibliography{dipole}

\begin{thebibliography}{73}%
\makeatletter
\providecommand \@ifxundefined [1]{%
 \@ifx{#1\undefined}
}%
\providecommand \@ifnum [1]{%
 \ifnum #1\expandafter \@firstoftwo
 \else \expandafter \@secondoftwo
 \fi
}%
\providecommand \@ifx [1]{%
 \ifx #1\expandafter \@firstoftwo
 \else \expandafter \@secondoftwo
 \fi
}%
\providecommand \natexlab [1]{#1}%
\providecommand \enquote  [1]{``#1''}%
\providecommand \bibnamefont  [1]{#1}%
\providecommand \bibfnamefont [1]{#1}%
\providecommand \citenamefont [1]{#1}%
\providecommand \href@noop [0]{\@secondoftwo}%
\providecommand \href [0]{\begingroup \@sanitize@url \@href}%
\providecommand \@href[1]{\@@startlink{#1}\@@href}%
\providecommand \@@href[1]{\endgroup#1\@@endlink}%
\providecommand \@sanitize@url [0]{\catcode `\\12\catcode `\$12\catcode
  `\&12\catcode `\#12\catcode `\^12\catcode `\_12\catcode `\%12\relax}%
\providecommand \@@startlink[1]{}%
\providecommand \@@endlink[0]{}%
\providecommand \url  [0]{\begingroup\@sanitize@url \@url }%
\providecommand \@url [1]{\endgroup\@href {#1}{\urlprefix }}%
\providecommand \urlprefix  [0]{URL }%
\providecommand \Eprint [0]{\href }%
\providecommand \doibase [0]{http://dx.doi.org/}%
\providecommand \selectlanguage [0]{\@gobble}%
\providecommand \bibinfo  [0]{\@secondoftwo}%
\providecommand \bibfield  [0]{\@secondoftwo}%
\providecommand \translation [1]{[#1]}%
\providecommand \BibitemOpen [0]{}%
\providecommand \bibitemStop [0]{}%
\providecommand \bibitemNoStop [0]{.\EOS\space}%
\providecommand \EOS [0]{\spacefactor3000\relax}%
\providecommand \BibitemShut  [1]{\csname bibitem#1\endcsname}%
\let\auto@bib@innerbib\@empty
\bibitem [{\citenamefont {Abbott}\ \emph
  {et~al.}(2019{\natexlab{a}})\citenamefont {Abbott} \emph
  {et~al.}}]{LIGOScientific:2018mvr}%
  \BibitemOpen
  \bibfield  {author} {\bibinfo {author} {\bibfnamefont {B.~P.}\ \bibnamefont
  {Abbott}} \emph {et~al.} (\bibinfo {collaboration} {LIGO, Virgo}),\
  }\bibfield  {title} {\enquote {\bibinfo {title} {{GWTC-1: A
  Gravitational-Wave Transient Catalog of Compact Binary Mergers Observed by
  LIGO and Virgo during the First and Second Observing Runs}},}\ }\href
  {\doibase 10.1103/PhysRevX.9.031040} {\bibfield  {journal} {\bibinfo
  {journal} {Phys. Rev. X}\ }\textbf {\bibinfo {volume} {9}},\ \bibinfo {pages}
  {031040} (\bibinfo {year} {2019}{\natexlab{a}})},\ \Eprint
  {http://arxiv.org/abs/1811.12907} {arXiv:1811.12907 [astro-ph.HE]}
  \BibitemShut {NoStop}%
\bibitem [{\citenamefont {Abbott}\ \emph
  {et~al.}(2021{\natexlab{a}})\citenamefont {Abbott} \emph
  {et~al.}}]{LIGOScientific:2020ibl}%
  \BibitemOpen
  \bibfield  {author} {\bibinfo {author} {\bibfnamefont {R.}~\bibnamefont
  {Abbott}} \emph {et~al.} (\bibinfo {collaboration} {LIGO, Virgo}),\
  }\bibfield  {title} {\enquote {\bibinfo {title} {{GWTC-2: Compact Binary
  Coalescences Observed by LIGO and Virgo During the First Half of the Third
  Observing Run}},}\ }\href {\doibase 10.1103/PhysRevX.11.021053} {\bibfield
  {journal} {\bibinfo  {journal} {Phys. Rev. X}\ }\textbf {\bibinfo {volume}
  {11}},\ \bibinfo {pages} {021053} (\bibinfo {year} {2021}{\natexlab{a}})},\
  \Eprint {http://arxiv.org/abs/2010.14527} {arXiv:2010.14527 [gr-qc]}
  \BibitemShut {NoStop}%
\bibitem [{\citenamefont {Abbott}\ \emph
  {et~al.}(2021{\natexlab{b}})\citenamefont {Abbott} \emph
  {et~al.}}]{LIGOScientific:2021usb}%
  \BibitemOpen
  \bibfield  {author} {\bibinfo {author} {\bibfnamefont {R.}~\bibnamefont
  {Abbott}} \emph {et~al.} (\bibinfo {collaboration} {LIGO, VIRGO}),\
  }\bibfield  {title} {\enquote {\bibinfo {title} {{GWTC-2.1: Deep Extended
  Catalog of Compact Binary Coalescences Observed by LIGO and Virgo During the
  First Half of the Third Observing Run}},}\ }\href@noop {} {\  (\bibinfo
  {year} {2021}{\natexlab{b}})},\ \Eprint {http://arxiv.org/abs/2108.01045}
  {arXiv:2108.01045 [gr-qc]} \BibitemShut {NoStop}%
\bibitem [{\citenamefont {Abbott}\ \emph
  {et~al.}(2021{\natexlab{c}})\citenamefont {Abbott} \emph
  {et~al.}}]{LIGOScientific:2021djp}%
  \BibitemOpen
  \bibfield  {author} {\bibinfo {author} {\bibfnamefont {R.}~\bibnamefont
  {Abbott}} \emph {et~al.} (\bibinfo {collaboration} {LIGO Scientific, VIRGO,
  KAGRA}),\ }\bibfield  {title} {\enquote {\bibinfo {title} {{GWTC-3: Compact
  Binary Coalescences Observed by LIGO and Virgo During the Second Part of the
  Third Observing Run}},}\ }\href@noop {} {\  (\bibinfo {year}
  {2021}{\natexlab{c}})},\ \Eprint {http://arxiv.org/abs/2111.03606}
  {arXiv:2111.03606 [gr-qc]} \BibitemShut {NoStop}%
\bibitem [{\citenamefont {Allen}(1996)}]{Allen:1996vm}%
  \BibitemOpen
  \bibfield  {author} {\bibinfo {author} {\bibfnamefont {Bruce}\ \bibnamefont
  {Allen}},\ }\bibfield  {title} {\enquote {\bibinfo {title} {{The Stochastic
  gravity wave background: Sources and detection}},}\ }in\ \href@noop {} {\emph
  {\bibinfo {booktitle} {{Les Houches School of Physics: Astrophysical Sources
  of Gravitational Radiation}}}}\ (\bibinfo {year} {1996})\ pp.\ \bibinfo
  {pages} {373--417},\ \Eprint {http://arxiv.org/abs/gr-qc/9604033}
  {arXiv:gr-qc/9604033} \BibitemShut {NoStop}%
\bibitem [{\citenamefont {Maggiore}(2000)}]{Maggiore:1999vm}%
  \BibitemOpen
  \bibfield  {author} {\bibinfo {author} {\bibfnamefont {Michele}\ \bibnamefont
  {Maggiore}},\ }\bibfield  {title} {\enquote {\bibinfo {title} {{Gravitational
  wave experiments and early universe cosmology}},}\ }\href {\doibase
  10.1016/S0370-1573(99)00102-7} {\bibfield  {journal} {\bibinfo  {journal}
  {Phys. Rept.}\ }\textbf {\bibinfo {volume} {331}},\ \bibinfo {pages}
  {283--367} (\bibinfo {year} {2000})},\ \Eprint
  {http://arxiv.org/abs/gr-qc/9909001} {arXiv:gr-qc/9909001} \BibitemShut
  {NoStop}%
\bibitem [{\citenamefont {Romano}\ and\ \citenamefont
  {Cornish}(2017)}]{Romano:2016dpx}%
  \BibitemOpen
  \bibfield  {author} {\bibinfo {author} {\bibfnamefont {Joseph~D.}\
  \bibnamefont {Romano}}\ and\ \bibinfo {author} {\bibfnamefont {Neil~J.}\
  \bibnamefont {Cornish}},\ }\bibfield  {title} {\enquote {\bibinfo {title}
  {{Detection methods for stochastic gravitational-wave backgrounds: a unified
  treatment}},}\ }\href {\doibase 10.1007/s41114-017-0004-1} {\bibfield
  {journal} {\bibinfo  {journal} {Living Rev. Rel.}\ }\textbf {\bibinfo
  {volume} {20}},\ \bibinfo {pages} {2} (\bibinfo {year} {2017})},\ \Eprint
  {http://arxiv.org/abs/1608.06889} {arXiv:1608.06889 [gr-qc]} \BibitemShut
  {NoStop}%
\bibitem [{\citenamefont {Christensen}(2019)}]{Christensen:2018iqi}%
  \BibitemOpen
  \bibfield  {author} {\bibinfo {author} {\bibfnamefont {Nelson}\ \bibnamefont
  {Christensen}},\ }\bibfield  {title} {\enquote {\bibinfo {title} {{Stochastic
  Gravitational Wave Backgrounds}},}\ }\href {\doibase
  10.1088/1361-6633/aae6b5} {\bibfield  {journal} {\bibinfo  {journal} {Rept.
  Prog. Phys.}\ }\textbf {\bibinfo {volume} {82}},\ \bibinfo {pages} {016903}
  (\bibinfo {year} {2019})},\ \Eprint {http://arxiv.org/abs/1811.08797}
  {arXiv:1811.08797 [gr-qc]} \BibitemShut {NoStop}%
\bibitem [{\citenamefont {Renzini}\ \emph {et~al.}(2022)\citenamefont
  {Renzini}, \citenamefont {Goncharov}, \citenamefont {Jenkins},\ and\
  \citenamefont {Meyers}}]{Renzini:2022alw}%
  \BibitemOpen
  \bibfield  {author} {\bibinfo {author} {\bibfnamefont {Arianna~I.}\
  \bibnamefont {Renzini}}, \bibinfo {author} {\bibfnamefont {Boris}\
  \bibnamefont {Goncharov}}, \bibinfo {author} {\bibfnamefont {Alexander~C.}\
  \bibnamefont {Jenkins}}, \ and\ \bibinfo {author} {\bibfnamefont {Pat~M.}\
  \bibnamefont {Meyers}},\ }\bibfield  {title} {\enquote {\bibinfo {title}
  {{Stochastic Gravitational-Wave Backgrounds: Current Detection Efforts and
  Future Prospects}},}\ }\href {\doibase 10.3390/galaxies10010034} {\bibfield
  {journal} {\bibinfo  {journal} {Galaxies}\ }\textbf {\bibinfo {volume}
  {10}},\ \bibinfo {pages} {34} (\bibinfo {year} {2022})},\ \Eprint
  {http://arxiv.org/abs/2202.00178} {arXiv:2202.00178 [gr-qc]} \BibitemShut
  {NoStop}%
\bibitem [{\citenamefont {Regimbau}(2011)}]{Regimbau:2011rp}%
  \BibitemOpen
  \bibfield  {author} {\bibinfo {author} {\bibfnamefont {Tania}\ \bibnamefont
  {Regimbau}},\ }\bibfield  {title} {\enquote {\bibinfo {title} {{The
  astrophysical gravitational wave stochastic background}},}\ }\href {\doibase
  10.1088/1674-4527/11/4/001} {\bibfield  {journal} {\bibinfo  {journal} {Res.
  Astron. Astrophys.}\ }\textbf {\bibinfo {volume} {11}},\ \bibinfo {pages}
  {369--390} (\bibinfo {year} {2011})},\ \Eprint
  {http://arxiv.org/abs/1101.2762} {arXiv:1101.2762 [astro-ph.CO]} \BibitemShut
  {NoStop}%
\bibitem [{\citenamefont {Martinovic}\ \emph
  {et~al.}(2021{\natexlab{a}})\citenamefont {Martinovic}, \citenamefont
  {Perigois}, \citenamefont {Regimbau},\ and\ \citenamefont
  {Sakellariadou}}]{Martinovic:2021fzj}%
  \BibitemOpen
  \bibfield  {author} {\bibinfo {author} {\bibfnamefont {Katarina}\
  \bibnamefont {Martinovic}}, \bibinfo {author} {\bibfnamefont {Carole}\
  \bibnamefont {Perigois}}, \bibinfo {author} {\bibfnamefont {Tania}\
  \bibnamefont {Regimbau}}, \ and\ \bibinfo {author} {\bibfnamefont {Mairi}\
  \bibnamefont {Sakellariadou}},\ }\bibfield  {title} {\enquote {\bibinfo
  {title} {{Footprints of population III stars in the gravitational-wave
  background}},}\ }\href@noop {} {\  (\bibinfo {year} {2021}{\natexlab{a}})},\
  \Eprint {http://arxiv.org/abs/2109.09779} {arXiv:2109.09779 [astro-ph.SR]}
  \BibitemShut {NoStop}%
\bibitem [{\citenamefont {Caprini}\ and\ \citenamefont
  {Figueroa}(2018)}]{Caprini:2018mtu}%
  \BibitemOpen
  \bibfield  {author} {\bibinfo {author} {\bibfnamefont {Chiara}\ \bibnamefont
  {Caprini}}\ and\ \bibinfo {author} {\bibfnamefont {Daniel~G.}\ \bibnamefont
  {Figueroa}},\ }\bibfield  {title} {\enquote {\bibinfo {title} {{Cosmological
  Backgrounds of Gravitational Waves}},}\ }\href {\doibase
  10.1088/1361-6382/aac608} {\bibfield  {journal} {\bibinfo  {journal} {Class.
  Quant. Grav.}\ }\textbf {\bibinfo {volume} {35}},\ \bibinfo {pages} {163001}
  (\bibinfo {year} {2018})},\ \Eprint {http://arxiv.org/abs/1801.04268}
  {arXiv:1801.04268 [astro-ph.CO]} \BibitemShut {NoStop}%
\bibitem [{\citenamefont {Abbott}\ \emph
  {et~al.}(2021{\natexlab{d}})\citenamefont {Abbott} \emph
  {et~al.}}]{LIGOScientific:2021nrg}%
  \BibitemOpen
  \bibfield  {author} {\bibinfo {author} {\bibfnamefont {R.}~\bibnamefont
  {Abbott}} \emph {et~al.} (\bibinfo {collaboration} {LIGO, Virgo, KAGRA}),\
  }\bibfield  {title} {\enquote {\bibinfo {title} {{Constraints on Cosmic
  Strings Using Data from the Third Advanced LIGO\textendash{}Virgo Observing
  Run}},}\ }\href {\doibase 10.1103/PhysRevLett.126.241102} {\bibfield
  {journal} {\bibinfo  {journal} {Phys. Rev. Lett.}\ }\textbf {\bibinfo
  {volume} {126}},\ \bibinfo {pages} {241102} (\bibinfo {year}
  {2021}{\natexlab{d}})},\ \Eprint {http://arxiv.org/abs/2101.12248}
  {arXiv:2101.12248 [gr-qc]} \BibitemShut {NoStop}%
\bibitem [{\citenamefont {Romero}\ \emph {et~al.}(2021)\citenamefont {Romero},
  \citenamefont {Martinovic}, \citenamefont {Callister}, \citenamefont {Guo},
  \citenamefont {Mart\'\i{}nez}, \citenamefont {Sakellariadou}, \citenamefont
  {Yang},\ and\ \citenamefont {Zhao}}]{Romero:2021kby}%
  \BibitemOpen
  \bibfield  {author} {\bibinfo {author} {\bibfnamefont {Alba}\ \bibnamefont
  {Romero}}, \bibinfo {author} {\bibfnamefont {Katarina}\ \bibnamefont
  {Martinovic}}, \bibinfo {author} {\bibfnamefont {Thomas~A.}\ \bibnamefont
  {Callister}}, \bibinfo {author} {\bibfnamefont {Huai-Ke}\ \bibnamefont
  {Guo}}, \bibinfo {author} {\bibfnamefont {Mario}\ \bibnamefont
  {Mart\'\i{}nez}}, \bibinfo {author} {\bibfnamefont {Mairi}\ \bibnamefont
  {Sakellariadou}}, \bibinfo {author} {\bibfnamefont {Feng-Wei}\ \bibnamefont
  {Yang}}, \ and\ \bibinfo {author} {\bibfnamefont {Yue}\ \bibnamefont
  {Zhao}},\ }\bibfield  {title} {\enquote {\bibinfo {title} {{Implications for
  First-Order Cosmological Phase Transitions from the Third LIGO-Virgo
  Observing Run}},}\ }\href {\doibase 10.1103/PhysRevLett.126.151301}
  {\bibfield  {journal} {\bibinfo  {journal} {Phys. Rev. Lett.}\ }\textbf
  {\bibinfo {volume} {126}},\ \bibinfo {pages} {151301} (\bibinfo {year}
  {2021})},\ \Eprint {http://arxiv.org/abs/2102.01714} {arXiv:2102.01714
  [hep-ph]} \BibitemShut {NoStop}%
\bibitem [{\citenamefont {Martinovic}\ \emph
  {et~al.}(2021{\natexlab{b}})\citenamefont {Martinovic}, \citenamefont
  {Badger}, \citenamefont {Sakellariadou},\ and\ \citenamefont
  {Mandic}}]{Martinovic:2021hzy}%
  \BibitemOpen
  \bibfield  {author} {\bibinfo {author} {\bibfnamefont {Katarina}\
  \bibnamefont {Martinovic}}, \bibinfo {author} {\bibfnamefont {Charles}\
  \bibnamefont {Badger}}, \bibinfo {author} {\bibfnamefont {Mairi}\
  \bibnamefont {Sakellariadou}}, \ and\ \bibinfo {author} {\bibfnamefont {Vuk}\
  \bibnamefont {Mandic}},\ }\bibfield  {title} {\enquote {\bibinfo {title}
  {{Searching for parity violation with the LIGO-Virgo-KAGRA network}},}\
  }\href {\doibase 10.1103/PhysRevD.104.L081101} {\bibfield  {journal}
  {\bibinfo  {journal} {Phys. Rev. D}\ }\textbf {\bibinfo {volume} {104}},\
  \bibinfo {pages} {L081101} (\bibinfo {year} {2021}{\natexlab{b}})},\ \Eprint
  {http://arxiv.org/abs/2103.06718} {arXiv:2103.06718 [gr-qc]} \BibitemShut
  {NoStop}%
\bibitem [{\citenamefont {Romero-Rodriguez}\ \emph {et~al.}(2022)\citenamefont
  {Romero-Rodriguez}, \citenamefont {Martinez}, \citenamefont {Pujol\`as},
  \citenamefont {Sakellariadou},\ and\ \citenamefont
  {Vaskonen}}]{Romero-Rodriguez:2021aws}%
  \BibitemOpen
  \bibfield  {author} {\bibinfo {author} {\bibfnamefont {Alba}\ \bibnamefont
  {Romero-Rodriguez}}, \bibinfo {author} {\bibfnamefont {Mario}\ \bibnamefont
  {Martinez}}, \bibinfo {author} {\bibfnamefont {Oriol}\ \bibnamefont
  {Pujol\`as}}, \bibinfo {author} {\bibfnamefont {Mairi}\ \bibnamefont
  {Sakellariadou}}, \ and\ \bibinfo {author} {\bibfnamefont {Ville}\
  \bibnamefont {Vaskonen}},\ }\bibfield  {title} {\enquote {\bibinfo {title}
  {{Search for a Scalar Induced Stochastic Gravitational Wave Background in the
  Third LIGO-Virgo Observing Run}},}\ }\href {\doibase
  10.1103/PhysRevLett.128.051301} {\bibfield  {journal} {\bibinfo  {journal}
  {Phys. Rev. Lett.}\ }\textbf {\bibinfo {volume} {128}},\ \bibinfo {pages}
  {051301} (\bibinfo {year} {2022})},\ \Eprint
  {http://arxiv.org/abs/2107.11660} {arXiv:2107.11660 [gr-qc]} \BibitemShut
  {NoStop}%
\bibitem [{\citenamefont {Sakellariadou}(2022)}]{Sakellariadou:2022tcm}%
  \BibitemOpen
  \bibfield  {author} {\bibinfo {author} {\bibfnamefont {Mairi}\ \bibnamefont
  {Sakellariadou}},\ }\bibfield  {title} {\enquote {\bibinfo {title}
  {{Gravitational Waves: The Theorist\textquoteright{}s Swiss Knife}},}\ }\href
  {\doibase 10.3390/universe8020132} {\bibfield  {journal} {\bibinfo  {journal}
  {Universe}\ }\textbf {\bibinfo {volume} {8}},\ \bibinfo {pages} {132}
  (\bibinfo {year} {2022})},\ \Eprint {http://arxiv.org/abs/2202.00735}
  {arXiv:2202.00735 [astro-ph.CO]} \BibitemShut {NoStop}%
\bibitem [{\citenamefont {Cusin}\ \emph {et~al.}(2017)\citenamefont {Cusin},
  \citenamefont {Pitrou},\ and\ \citenamefont {Uzan}}]{Cusin:2017fwz}%
  \BibitemOpen
  \bibfield  {author} {\bibinfo {author} {\bibfnamefont {Giulia}\ \bibnamefont
  {Cusin}}, \bibinfo {author} {\bibfnamefont {Cyril}\ \bibnamefont {Pitrou}}, \
  and\ \bibinfo {author} {\bibfnamefont {Jean-Philippe}\ \bibnamefont {Uzan}},\
  }\bibfield  {title} {\enquote {\bibinfo {title} {{Anisotropy of the
  astrophysical gravitational wave background: Analytic expression of the
  angular power spectrum and correlation with cosmological observations}},}\
  }\href {\doibase 10.1103/PhysRevD.96.103019} {\bibfield  {journal} {\bibinfo
  {journal} {Phys. Rev. D}\ }\textbf {\bibinfo {volume} {96}},\ \bibinfo
  {pages} {103019} (\bibinfo {year} {2017})},\ \Eprint
  {http://arxiv.org/abs/1704.06184} {arXiv:1704.06184 [astro-ph.CO]}
  \BibitemShut {NoStop}%
\bibitem [{\citenamefont {Cusin}\ \emph
  {et~al.}(2018{\natexlab{a}})\citenamefont {Cusin}, \citenamefont {Pitrou},\
  and\ \citenamefont {Uzan}}]{Cusin:2017mjm}%
  \BibitemOpen
  \bibfield  {author} {\bibinfo {author} {\bibfnamefont {Giulia}\ \bibnamefont
  {Cusin}}, \bibinfo {author} {\bibfnamefont {Cyril}\ \bibnamefont {Pitrou}}, \
  and\ \bibinfo {author} {\bibfnamefont {Jean-Philippe}\ \bibnamefont {Uzan}},\
  }\bibfield  {title} {\enquote {\bibinfo {title} {{The signal of the
  gravitational wave background and the angular correlation of its energy
  density}},}\ }\href {\doibase 10.1103/PhysRevD.97.123527} {\bibfield
  {journal} {\bibinfo  {journal} {Phys. Rev. D}\ }\textbf {\bibinfo {volume}
  {97}},\ \bibinfo {pages} {123527} (\bibinfo {year} {2018}{\natexlab{a}})},\
  \Eprint {http://arxiv.org/abs/1711.11345} {arXiv:1711.11345 [astro-ph.CO]}
  \BibitemShut {NoStop}%
\bibitem [{\citenamefont {Jenkins}\ and\ \citenamefont
  {Sakellariadou}(2018)}]{Jenkins:2018nty}%
  \BibitemOpen
  \bibfield  {author} {\bibinfo {author} {\bibfnamefont {Alexander~C.}\
  \bibnamefont {Jenkins}}\ and\ \bibinfo {author} {\bibfnamefont {Mairi}\
  \bibnamefont {Sakellariadou}},\ }\bibfield  {title} {\enquote {\bibinfo
  {title} {{Anisotropies in the stochastic gravitational-wave background:
  Formalism and the cosmic string case}},}\ }\href {\doibase
  10.1103/PhysRevD.98.063509} {\bibfield  {journal} {\bibinfo  {journal} {Phys.
  Rev. D}\ }\textbf {\bibinfo {volume} {98}},\ \bibinfo {pages} {063509}
  (\bibinfo {year} {2018})},\ \Eprint {http://arxiv.org/abs/1802.06046}
  {arXiv:1802.06046 [astro-ph.CO]} \BibitemShut {NoStop}%
\bibitem [{\citenamefont {Cusin}\ \emph
  {et~al.}(2018{\natexlab{b}})\citenamefont {Cusin}, \citenamefont {Dvorkin},
  \citenamefont {Pitrou},\ and\ \citenamefont {Uzan}}]{Cusin:2018rsq}%
  \BibitemOpen
  \bibfield  {author} {\bibinfo {author} {\bibfnamefont {Giulia}\ \bibnamefont
  {Cusin}}, \bibinfo {author} {\bibfnamefont {Irina}\ \bibnamefont {Dvorkin}},
  \bibinfo {author} {\bibfnamefont {Cyril}\ \bibnamefont {Pitrou}}, \ and\
  \bibinfo {author} {\bibfnamefont {Jean-Philippe}\ \bibnamefont {Uzan}},\
  }\bibfield  {title} {\enquote {\bibinfo {title} {{First predictions of the
  angular power spectrum of the astrophysical gravitational wave
  background}},}\ }\href {\doibase 10.1103/PhysRevLett.120.231101} {\bibfield
  {journal} {\bibinfo  {journal} {Phys. Rev. Lett.}\ }\textbf {\bibinfo
  {volume} {120}},\ \bibinfo {pages} {231101} (\bibinfo {year}
  {2018}{\natexlab{b}})},\ \Eprint {http://arxiv.org/abs/1803.03236}
  {arXiv:1803.03236 [astro-ph.CO]} \BibitemShut {NoStop}%
\bibitem [{\citenamefont {Geller}\ \emph {et~al.}(2018)\citenamefont {Geller},
  \citenamefont {Hook}, \citenamefont {Sundrum},\ and\ \citenamefont
  {Tsai}}]{Geller:2018mwu}%
  \BibitemOpen
  \bibfield  {author} {\bibinfo {author} {\bibfnamefont {Michael}\ \bibnamefont
  {Geller}}, \bibinfo {author} {\bibfnamefont {Anson}\ \bibnamefont {Hook}},
  \bibinfo {author} {\bibfnamefont {Raman}\ \bibnamefont {Sundrum}}, \ and\
  \bibinfo {author} {\bibfnamefont {Yuhsin}\ \bibnamefont {Tsai}},\ }\bibfield
  {title} {\enquote {\bibinfo {title} {{Primordial Anisotropies in the
  Gravitational Wave Background from Cosmological Phase Transitions}},}\ }\href
  {\doibase 10.1103/PhysRevLett.121.201303} {\bibfield  {journal} {\bibinfo
  {journal} {Phys. Rev. Lett.}\ }\textbf {\bibinfo {volume} {121}},\ \bibinfo
  {pages} {201303} (\bibinfo {year} {2018})},\ \Eprint
  {http://arxiv.org/abs/1803.10780} {arXiv:1803.10780 [hep-ph]} \BibitemShut
  {NoStop}%
\bibitem [{\citenamefont {Jenkins}\ \emph {et~al.}(2018)\citenamefont
  {Jenkins}, \citenamefont {Sakellariadou}, \citenamefont {Regimbau},\ and\
  \citenamefont {Slezak}}]{Jenkins:2018uac}%
  \BibitemOpen
  \bibfield  {author} {\bibinfo {author} {\bibfnamefont {Alexander~C.}\
  \bibnamefont {Jenkins}}, \bibinfo {author} {\bibfnamefont {Mairi}\
  \bibnamefont {Sakellariadou}}, \bibinfo {author} {\bibfnamefont {Tania}\
  \bibnamefont {Regimbau}}, \ and\ \bibinfo {author} {\bibfnamefont {Eric}\
  \bibnamefont {Slezak}},\ }\bibfield  {title} {\enquote {\bibinfo {title}
  {{Anisotropies in the astrophysical gravitational-wave background:
  Predictions for the detection of compact binaries by LIGO and Virgo}},}\
  }\href {\doibase 10.1103/PhysRevD.98.063501} {\bibfield  {journal} {\bibinfo
  {journal} {Phys. Rev. D}\ }\textbf {\bibinfo {volume} {98}},\ \bibinfo
  {pages} {063501} (\bibinfo {year} {2018})},\ \Eprint
  {http://arxiv.org/abs/1806.01718} {arXiv:1806.01718 [astro-ph.CO]}
  \BibitemShut {NoStop}%
\bibitem [{\citenamefont {Jenkins}\ \emph
  {et~al.}(2019{\natexlab{a}})\citenamefont {Jenkins}, \citenamefont
  {O'Shaughnessy}, \citenamefont {Sakellariadou},\ and\ \citenamefont
  {Wysocki}}]{Jenkins:2018kxc}%
  \BibitemOpen
  \bibfield  {author} {\bibinfo {author} {\bibfnamefont {Alexander~C.}\
  \bibnamefont {Jenkins}}, \bibinfo {author} {\bibfnamefont {Richard}\
  \bibnamefont {O'Shaughnessy}}, \bibinfo {author} {\bibfnamefont {Mairi}\
  \bibnamefont {Sakellariadou}}, \ and\ \bibinfo {author} {\bibfnamefont
  {Daniel}\ \bibnamefont {Wysocki}},\ }\bibfield  {title} {\enquote {\bibinfo
  {title} {{Anisotropies in the astrophysical gravitational-wave background:
  The impact of black hole distributions}},}\ }\href {\doibase
  10.1103/PhysRevLett.122.111101} {\bibfield  {journal} {\bibinfo  {journal}
  {Phys. Rev. Lett.}\ }\textbf {\bibinfo {volume} {122}},\ \bibinfo {pages}
  {111101} (\bibinfo {year} {2019}{\natexlab{a}})},\ \Eprint
  {http://arxiv.org/abs/1810.13435} {arXiv:1810.13435 [astro-ph.CO]}
  \BibitemShut {NoStop}%
\bibitem [{\citenamefont {Bartolo}\ \emph
  {et~al.}(2020{\natexlab{a}})\citenamefont {Bartolo}, \citenamefont
  {Bertacca}, \citenamefont {De~Luca}, \citenamefont {Franciolini},
  \citenamefont {Matarrese}, \citenamefont {Peloso}, \citenamefont
  {Ricciardone}, \citenamefont {Riotto},\ and\ \citenamefont
  {Tasinato}}]{Bartolo:2019zvb}%
  \BibitemOpen
  \bibfield  {author} {\bibinfo {author} {\bibfnamefont {N.}~\bibnamefont
  {Bartolo}}, \bibinfo {author} {\bibfnamefont {D.}~\bibnamefont {Bertacca}},
  \bibinfo {author} {\bibfnamefont {V.}~\bibnamefont {De~Luca}}, \bibinfo
  {author} {\bibfnamefont {G.}~\bibnamefont {Franciolini}}, \bibinfo {author}
  {\bibfnamefont {S.}~\bibnamefont {Matarrese}}, \bibinfo {author}
  {\bibfnamefont {M.}~\bibnamefont {Peloso}}, \bibinfo {author} {\bibfnamefont
  {A.}~\bibnamefont {Ricciardone}}, \bibinfo {author} {\bibfnamefont
  {A.}~\bibnamefont {Riotto}}, \ and\ \bibinfo {author} {\bibfnamefont
  {G.}~\bibnamefont {Tasinato}},\ }\bibfield  {title} {\enquote {\bibinfo
  {title} {{Gravitational wave anisotropies from primordial black holes}},}\
  }\href {\doibase 10.1088/1475-7516/2020/02/028} {\bibfield  {journal}
  {\bibinfo  {journal} {JCAP}\ }\textbf {\bibinfo {volume} {02}},\ \bibinfo
  {pages} {028} (\bibinfo {year} {2020}{\natexlab{a}})},\ \Eprint
  {http://arxiv.org/abs/1909.12619} {arXiv:1909.12619 [astro-ph.CO]}
  \BibitemShut {NoStop}%
\bibitem [{\citenamefont {Bertacca}\ \emph {et~al.}(2020)\citenamefont
  {Bertacca}, \citenamefont {Ricciardone}, \citenamefont {Bellomo},
  \citenamefont {Jenkins}, \citenamefont {Matarrese}, \citenamefont
  {Raccanelli}, \citenamefont {Regimbau},\ and\ \citenamefont
  {Sakellariadou}}]{Bertacca:2019fnt}%
  \BibitemOpen
  \bibfield  {author} {\bibinfo {author} {\bibfnamefont {Daniele}\ \bibnamefont
  {Bertacca}}, \bibinfo {author} {\bibfnamefont {Angelo}\ \bibnamefont
  {Ricciardone}}, \bibinfo {author} {\bibfnamefont {Nicola}\ \bibnamefont
  {Bellomo}}, \bibinfo {author} {\bibfnamefont {Alexander~C.}\ \bibnamefont
  {Jenkins}}, \bibinfo {author} {\bibfnamefont {Sabino}\ \bibnamefont
  {Matarrese}}, \bibinfo {author} {\bibfnamefont {Alvise}\ \bibnamefont
  {Raccanelli}}, \bibinfo {author} {\bibfnamefont {Tania}\ \bibnamefont
  {Regimbau}}, \ and\ \bibinfo {author} {\bibfnamefont {Mairi}\ \bibnamefont
  {Sakellariadou}},\ }\bibfield  {title} {\enquote {\bibinfo {title}
  {{Projection effects on the observed angular spectrum of the astrophysical
  stochastic gravitational wave background}},}\ }\href {\doibase
  10.1103/PhysRevD.101.103513} {\bibfield  {journal} {\bibinfo  {journal}
  {Phys. Rev. D}\ }\textbf {\bibinfo {volume} {101}},\ \bibinfo {pages}
  {103513} (\bibinfo {year} {2020})},\ \Eprint
  {http://arxiv.org/abs/1909.11627} {arXiv:1909.11627 [astro-ph.CO]}
  \BibitemShut {NoStop}%
\bibitem [{\citenamefont {Bartolo}\ \emph {et~al.}(2019)\citenamefont
  {Bartolo}, \citenamefont {Bertacca}, \citenamefont {Matarrese}, \citenamefont
  {Peloso}, \citenamefont {Ricciardone}, \citenamefont {Riotto},\ and\
  \citenamefont {Tasinato}}]{Bartolo:2019oiq}%
  \BibitemOpen
  \bibfield  {author} {\bibinfo {author} {\bibfnamefont {N.}~\bibnamefont
  {Bartolo}}, \bibinfo {author} {\bibfnamefont {D.}~\bibnamefont {Bertacca}},
  \bibinfo {author} {\bibfnamefont {S.}~\bibnamefont {Matarrese}}, \bibinfo
  {author} {\bibfnamefont {M.}~\bibnamefont {Peloso}}, \bibinfo {author}
  {\bibfnamefont {A.}~\bibnamefont {Ricciardone}}, \bibinfo {author}
  {\bibfnamefont {A.}~\bibnamefont {Riotto}}, \ and\ \bibinfo {author}
  {\bibfnamefont {G.}~\bibnamefont {Tasinato}},\ }\bibfield  {title} {\enquote
  {\bibinfo {title} {{Anisotropies and non-Gaussianity of the Cosmological
  Gravitational Wave Background}},}\ }\href {\doibase
  10.1103/PhysRevD.100.121501} {\bibfield  {journal} {\bibinfo  {journal}
  {Phys. Rev. D}\ }\textbf {\bibinfo {volume} {100}},\ \bibinfo {pages}
  {121501} (\bibinfo {year} {2019})},\ \Eprint
  {http://arxiv.org/abs/1908.00527} {arXiv:1908.00527 [astro-ph.CO]}
  \BibitemShut {NoStop}%
\bibitem [{\citenamefont {Bartolo}\ \emph
  {et~al.}(2020{\natexlab{b}})\citenamefont {Bartolo}, \citenamefont
  {Bertacca}, \citenamefont {Matarrese}, \citenamefont {Peloso}, \citenamefont
  {Ricciardone}, \citenamefont {Riotto},\ and\ \citenamefont
  {Tasinato}}]{Bartolo:2019yeu}%
  \BibitemOpen
  \bibfield  {author} {\bibinfo {author} {\bibfnamefont {Nicola}\ \bibnamefont
  {Bartolo}}, \bibinfo {author} {\bibfnamefont {Daniele}\ \bibnamefont
  {Bertacca}}, \bibinfo {author} {\bibfnamefont {Sabino}\ \bibnamefont
  {Matarrese}}, \bibinfo {author} {\bibfnamefont {Marco}\ \bibnamefont
  {Peloso}}, \bibinfo {author} {\bibfnamefont {Angelo}\ \bibnamefont
  {Ricciardone}}, \bibinfo {author} {\bibfnamefont {Antonio}\ \bibnamefont
  {Riotto}}, \ and\ \bibinfo {author} {\bibfnamefont {Gianmassimo}\
  \bibnamefont {Tasinato}},\ }\bibfield  {title} {\enquote {\bibinfo {title}
  {{Characterizing the cosmological gravitational wave background: Anisotropies
  and non-Gaussianity}},}\ }\href {\doibase 10.1103/PhysRevD.102.023527}
  {\bibfield  {journal} {\bibinfo  {journal} {Phys. Rev. D}\ }\textbf {\bibinfo
  {volume} {102}},\ \bibinfo {pages} {023527} (\bibinfo {year}
  {2020}{\natexlab{b}})},\ \Eprint {http://arxiv.org/abs/1912.09433}
  {arXiv:1912.09433 [astro-ph.CO]} \BibitemShut {NoStop}%
\bibitem [{\citenamefont {Bellomo}\ \emph {et~al.}(2021)\citenamefont
  {Bellomo}, \citenamefont {Bertacca}, \citenamefont {Jenkins}, \citenamefont
  {Matarrese}, \citenamefont {Raccanelli}, \citenamefont {Regimbau},
  \citenamefont {Ricciardone},\ and\ \citenamefont
  {Sakellariadou}}]{Bellomo:2021mer}%
  \BibitemOpen
  \bibfield  {author} {\bibinfo {author} {\bibfnamefont {Nicola}\ \bibnamefont
  {Bellomo}}, \bibinfo {author} {\bibfnamefont {Daniele}\ \bibnamefont
  {Bertacca}}, \bibinfo {author} {\bibfnamefont {Alexander~C.}\ \bibnamefont
  {Jenkins}}, \bibinfo {author} {\bibfnamefont {Sabino}\ \bibnamefont
  {Matarrese}}, \bibinfo {author} {\bibfnamefont {Alvise}\ \bibnamefont
  {Raccanelli}}, \bibinfo {author} {\bibfnamefont {Tania}\ \bibnamefont
  {Regimbau}}, \bibinfo {author} {\bibfnamefont {Angelo}\ \bibnamefont
  {Ricciardone}}, \ and\ \bibinfo {author} {\bibfnamefont {Mairi}\ \bibnamefont
  {Sakellariadou}},\ }\bibfield  {title} {\enquote {\bibinfo {title}
  {{CLASS\_GWB: robust modeling of the astrophysical gravitational wave
  background anisotropies}},}\ }\href@noop {} {\  (\bibinfo {year} {2021})},\
  \Eprint {http://arxiv.org/abs/2110.15059} {arXiv:2110.15059 [gr-qc]}
  \BibitemShut {NoStop}%
\bibitem [{\citenamefont {Bartolo}\ \emph {et~al.}(2022)\citenamefont {Bartolo}
  \emph {et~al.}}]{LISACosmologyWorkingGroup:2022kbp}%
  \BibitemOpen
  \bibfield  {author} {\bibinfo {author} {\bibfnamefont {Nicola}\ \bibnamefont
  {Bartolo}} \emph {et~al.} (\bibinfo {collaboration} {LISA Cosmology Working
  Group}),\ }\bibfield  {title} {\enquote {\bibinfo {title} {{Probing
  Anisotropies of the Stochastic Gravitational Wave Background with LISA}},}\
  }\href@noop {} {\  (\bibinfo {year} {2022})},\ \Eprint
  {http://arxiv.org/abs/2201.08782} {arXiv:2201.08782 [astro-ph.CO]}
  \BibitemShut {NoStop}%
\bibitem [{\citenamefont {Jenkins}\ and\ \citenamefont
  {Sakellariadou}(2019)}]{Jenkins:2019uzp}%
  \BibitemOpen
  \bibfield  {author} {\bibinfo {author} {\bibfnamefont {Alexander~C.}\
  \bibnamefont {Jenkins}}\ and\ \bibinfo {author} {\bibfnamefont {Mairi}\
  \bibnamefont {Sakellariadou}},\ }\bibfield  {title} {\enquote {\bibinfo
  {title} {{Shot noise in the astrophysical gravitational-wave background}},}\
  }\href {\doibase 10.1103/PhysRevD.100.063508} {\bibfield  {journal} {\bibinfo
   {journal} {Phys. Rev. D}\ }\textbf {\bibinfo {volume} {100}},\ \bibinfo
  {pages} {063508} (\bibinfo {year} {2019})},\ \Eprint
  {http://arxiv.org/abs/1902.07719} {arXiv:1902.07719 [astro-ph.CO]}
  \BibitemShut {NoStop}%
\bibitem [{\citenamefont {Jenkins}\ \emph
  {et~al.}(2019{\natexlab{b}})\citenamefont {Jenkins}, \citenamefont {Romano},\
  and\ \citenamefont {Sakellariadou}}]{Jenkins:2019nks}%
  \BibitemOpen
  \bibfield  {author} {\bibinfo {author} {\bibfnamefont {Alexander~C.}\
  \bibnamefont {Jenkins}}, \bibinfo {author} {\bibfnamefont {Joseph~D.}\
  \bibnamefont {Romano}}, \ and\ \bibinfo {author} {\bibfnamefont {Mairi}\
  \bibnamefont {Sakellariadou}},\ }\bibfield  {title} {\enquote {\bibinfo
  {title} {{Estimating the angular power spectrum of the gravitational-wave
  background in the presence of shot noise}},}\ }\href {\doibase
  10.1103/PhysRevD.100.083501} {\bibfield  {journal} {\bibinfo  {journal}
  {Phys. Rev. D}\ }\textbf {\bibinfo {volume} {100}},\ \bibinfo {pages}
  {083501} (\bibinfo {year} {2019}{\natexlab{b}})},\ \Eprint
  {http://arxiv.org/abs/1907.06642} {arXiv:1907.06642 [astro-ph.CO]}
  \BibitemShut {NoStop}%
\bibitem [{\citenamefont {Cusin}\ and\ \citenamefont
  {Tasinato}(2022)}]{Cusin:2022cbb}%
  \BibitemOpen
  \bibfield  {author} {\bibinfo {author} {\bibfnamefont {Giulia}\ \bibnamefont
  {Cusin}}\ and\ \bibinfo {author} {\bibfnamefont {Gianmassimo}\ \bibnamefont
  {Tasinato}},\ }\bibfield  {title} {\enquote {\bibinfo {title} {{Doppler
  boosting the stochastic gravitational wave background}},}\ }\href@noop {} {\
  (\bibinfo {year} {2022})},\ \Eprint {http://arxiv.org/abs/2201.10464}
  {arXiv:2201.10464 [astro-ph.CO]} \BibitemShut {NoStop}%
\bibitem [{\citenamefont {Jenkins}(2022)}]{Jenkins:2022chv}%
  \BibitemOpen
  \bibfield  {author} {\bibinfo {author} {\bibfnamefont {Alexander~C.}\
  \bibnamefont {Jenkins}},\ }\emph {\bibinfo {title} {{Cosmology and
  Fundamental Physics in the Era of Gravitational-Wave Astronomy}}},\ \href
  {https://kclpure.kcl.ac.uk/portal/en/theses/cosmology-and-fundamental-physics-in-the-era-of-gravitationalwave-astronomy(39b9da34-05c4-4887-afe8-7bdcf4cad4c4).html}
  {Ph.D. thesis},\ \bibinfo  {school} {King's Coll. London} (\bibinfo {year}
  {2022}),\ \Eprint {http://arxiv.org/abs/2202.05105} {arXiv:2202.05105
  [gr-qc]} \BibitemShut {NoStop}%
\bibitem [{\citenamefont {Dall'Armi}\ \emph {et~al.}(2022)\citenamefont
  {Dall'Armi}, \citenamefont {Ricciardone},\ and\ \citenamefont
  {Bertacca}}]{DallArmi:2022wnq}%
  \BibitemOpen
  \bibfield  {author} {\bibinfo {author} {\bibfnamefont {Lorenzo~Valbusa}\
  \bibnamefont {Dall'Armi}}, \bibinfo {author} {\bibfnamefont {Angelo}\
  \bibnamefont {Ricciardone}}, \ and\ \bibinfo {author} {\bibfnamefont
  {Daniele}\ \bibnamefont {Bertacca}},\ }\bibfield  {title} {\enquote {\bibinfo
  {title} {{The Dipole of the Astrophysical Gravitational-Wave Background}},}\
  }\href@noop {} {\  (\bibinfo {year} {2022})},\ \Eprint
  {http://arxiv.org/abs/2206.02747} {arXiv:2206.02747 [astro-ph.CO]}
  \BibitemShut {NoStop}%
\bibitem [{\citenamefont {Aghanim}\ \emph {et~al.}(2020)\citenamefont {Aghanim}
  \emph {et~al.}}]{Planck:2018nkj}%
  \BibitemOpen
  \bibfield  {author} {\bibinfo {author} {\bibfnamefont {N.}~\bibnamefont
  {Aghanim}} \emph {et~al.} (\bibinfo {collaboration} {Planck}),\ }\bibfield
  {title} {\enquote {\bibinfo {title} {{Planck 2018 results. I. Overview and
  the cosmological legacy of Planck}},}\ }\href {\doibase
  10.1051/0004-6361/201833880} {\bibfield  {journal} {\bibinfo  {journal}
  {Astron. Astrophys.}\ }\textbf {\bibinfo {volume} {641}},\ \bibinfo {pages}
  {A1} (\bibinfo {year} {2020})},\ \Eprint {http://arxiv.org/abs/1807.06205}
  {arXiv:1807.06205 [astro-ph.CO]} \BibitemShut {NoStop}%
\bibitem [{\citenamefont {Secrest}\ \emph {et~al.}(2021)\citenamefont
  {Secrest}, \citenamefont {von Hausegger}, \citenamefont {Rameez},
  \citenamefont {Mohayaee}, \citenamefont {Sarkar},\ and\ \citenamefont
  {Colin}}]{Secrest:2020has}%
  \BibitemOpen
  \bibfield  {author} {\bibinfo {author} {\bibfnamefont {Nathan~J.}\
  \bibnamefont {Secrest}}, \bibinfo {author} {\bibfnamefont {Sebastian}\
  \bibnamefont {von Hausegger}}, \bibinfo {author} {\bibfnamefont {Mohamed}\
  \bibnamefont {Rameez}}, \bibinfo {author} {\bibfnamefont {Roya}\ \bibnamefont
  {Mohayaee}}, \bibinfo {author} {\bibfnamefont {Subir}\ \bibnamefont
  {Sarkar}}, \ and\ \bibinfo {author} {\bibfnamefont {Jacques}\ \bibnamefont
  {Colin}},\ }\bibfield  {title} {\enquote {\bibinfo {title} {{A Test of the
  Cosmological Principle with Quasars}},}\ }\href {\doibase
  10.3847/2041-8213/abdd40} {\bibfield  {journal} {\bibinfo  {journal}
  {Astrophys. J. Lett.}\ }\textbf {\bibinfo {volume} {908}},\ \bibinfo {pages}
  {L51} (\bibinfo {year} {2021})},\ \Eprint {http://arxiv.org/abs/2009.14826}
  {arXiv:2009.14826 [astro-ph.CO]} \BibitemShut {NoStop}%
\bibitem [{\citenamefont {Dalang}\ and\ \citenamefont
  {Bonvin}(2022)}]{Dalang:2021ruy}%
  \BibitemOpen
  \bibfield  {author} {\bibinfo {author} {\bibfnamefont {Charles}\ \bibnamefont
  {Dalang}}\ and\ \bibinfo {author} {\bibfnamefont {Camille}\ \bibnamefont
  {Bonvin}},\ }\bibfield  {title} {\enquote {\bibinfo {title} {{On the
  kinematic cosmic dipole tension}},}\ }\href {\doibase 10.1093/mnras/stac726}
  {\bibfield  {journal} {\bibinfo  {journal} {Mon. Not. Roy. Astron. Soc.}\
  }\textbf {\bibinfo {volume} {512}},\ \bibinfo {pages} {3895--3905} (\bibinfo
  {year} {2022})},\ \Eprint {http://arxiv.org/abs/2111.03616} {arXiv:2111.03616
  [astro-ph.CO]} \BibitemShut {NoStop}%
\bibitem [{\citenamefont {Secrest}\ \emph {et~al.}(2022)\citenamefont
  {Secrest}, \citenamefont {von Hausegger}, \citenamefont {Rameez},
  \citenamefont {Mohayaee},\ and\ \citenamefont {Sarkar}}]{Secrest:2022uvx}%
  \BibitemOpen
  \bibfield  {author} {\bibinfo {author} {\bibfnamefont {Nathan}\ \bibnamefont
  {Secrest}}, \bibinfo {author} {\bibfnamefont {Sebastian}\ \bibnamefont {von
  Hausegger}}, \bibinfo {author} {\bibfnamefont {Mohamed}\ \bibnamefont
  {Rameez}}, \bibinfo {author} {\bibfnamefont {Roya}\ \bibnamefont {Mohayaee}},
  \ and\ \bibinfo {author} {\bibfnamefont {Subir}\ \bibnamefont {Sarkar}},\
  }\bibfield  {title} {\enquote {\bibinfo {title} {{A Challenge to the Standard
  Cosmological Model}},}\ }\href@noop {} {\  (\bibinfo {year} {2022})},\
  \Eprint {http://arxiv.org/abs/2206.05624} {arXiv:2206.05624 [astro-ph.CO]}
  \BibitemShut {NoStop}%
\bibitem [{\citenamefont {Darling}(2022)}]{Darling:2022jxt}%
  \BibitemOpen
  \bibfield  {author} {\bibinfo {author} {\bibfnamefont {Jeremy}\ \bibnamefont
  {Darling}},\ }\bibfield  {title} {\enquote {\bibinfo {title} {{The Universe
  is Brighter in the Direction of Our Motion: Galaxy Counts and Fluxes are
  Consistent with the CMB Dipole}},}\ }\href {\doibase
  10.3847/2041-8213/ac6f08} {\bibfield  {journal} {\bibinfo  {journal}
  {Astrophys. J. Lett.}\ }\textbf {\bibinfo {volume} {931}},\ \bibinfo {pages}
  {L14} (\bibinfo {year} {2022})},\ \Eprint {http://arxiv.org/abs/2205.06880}
  {arXiv:2205.06880 [astro-ph.CO]} \BibitemShut {NoStop}%
\bibitem [{\citenamefont {Abbott}\ \emph
  {et~al.}(2021{\natexlab{e}})\citenamefont {Abbott} \emph
  {et~al.}}]{KAGRA:2021kbb}%
  \BibitemOpen
  \bibfield  {author} {\bibinfo {author} {\bibfnamefont {R.}~\bibnamefont
  {Abbott}} \emph {et~al.} (\bibinfo {collaboration} {LIGO, Virgo, KAGRA}),\
  }\bibfield  {title} {\enquote {\bibinfo {title} {{Upper limits on the
  isotropic gravitational-wave background from Advanced LIGO and Advanced
  Virgo\textquoteright{}s third observing run}},}\ }\href {\doibase
  10.1103/PhysRevD.104.022004} {\bibfield  {journal} {\bibinfo  {journal}
  {Phys. Rev. D}\ }\textbf {\bibinfo {volume} {104}},\ \bibinfo {pages}
  {022004} (\bibinfo {year} {2021}{\natexlab{e}})},\ \Eprint
  {http://arxiv.org/abs/2101.12130} {arXiv:2101.12130 [gr-qc]} \BibitemShut
  {NoStop}%
\bibitem [{\citenamefont {Abbott}\ \emph
  {et~al.}(2017{\natexlab{a}})\citenamefont {Abbott} \emph
  {et~al.}}]{LIGOScientific:2016jlg}%
  \BibitemOpen
  \bibfield  {author} {\bibinfo {author} {\bibfnamefont {Benjamin~P.}\
  \bibnamefont {Abbott}} \emph {et~al.} (\bibinfo {collaboration} {LIGO,
  Virgo}),\ }\bibfield  {title} {\enquote {\bibinfo {title} {{Upper Limits on
  the Stochastic Gravitational-Wave Background from Advanced
  LIGO\textquoteright{}s First Observing Run}},}\ }\href {\doibase
  10.1103/PhysRevLett.118.121101} {\bibfield  {journal} {\bibinfo  {journal}
  {Phys. Rev. Lett.}\ }\textbf {\bibinfo {volume} {118}},\ \bibinfo {pages}
  {121101} (\bibinfo {year} {2017}{\natexlab{a}})},\ \bibinfo {note} {[Erratum:
  Phys.Rev.Lett. 119, 029901 (2017)]},\ \Eprint
  {http://arxiv.org/abs/1612.02029} {arXiv:1612.02029 [gr-qc]} \BibitemShut
  {NoStop}%
\bibitem [{\citenamefont {Abbott}\ \emph
  {et~al.}(2019{\natexlab{b}})\citenamefont {Abbott} \emph
  {et~al.}}]{LIGOScientific:2019vic}%
  \BibitemOpen
  \bibfield  {author} {\bibinfo {author} {\bibfnamefont {B.~P.}\ \bibnamefont
  {Abbott}} \emph {et~al.} (\bibinfo {collaboration} {LIGO, Virgo}),\
  }\bibfield  {title} {\enquote {\bibinfo {title} {{Search for the isotropic
  stochastic background using data from Advanced LIGO\textquoteright{}s second
  observing run}},}\ }\href {\doibase 10.1103/PhysRevD.100.061101} {\bibfield
  {journal} {\bibinfo  {journal} {Phys. Rev. D}\ }\textbf {\bibinfo {volume}
  {100}},\ \bibinfo {pages} {061101} (\bibinfo {year} {2019}{\natexlab{b}})},\
  \Eprint {http://arxiv.org/abs/1903.02886} {arXiv:1903.02886 [gr-qc]}
  \BibitemShut {NoStop}%
\bibitem [{\citenamefont {Abbott}\ \emph
  {et~al.}(2017{\natexlab{b}})\citenamefont {Abbott} \emph
  {et~al.}}]{LIGOScientific:2016nwa}%
  \BibitemOpen
  \bibfield  {author} {\bibinfo {author} {\bibfnamefont {Benjamin~P.}\
  \bibnamefont {Abbott}} \emph {et~al.} (\bibinfo {collaboration} {LIGO,
  Virgo}),\ }\bibfield  {title} {\enquote {\bibinfo {title} {{Directional
  Limits on Persistent Gravitational Waves from Advanced LIGO\textquoteright{}s
  First Observing Run}},}\ }\href {\doibase 10.1103/PhysRevLett.118.121102}
  {\bibfield  {journal} {\bibinfo  {journal} {Phys. Rev. Lett.}\ }\textbf
  {\bibinfo {volume} {118}},\ \bibinfo {pages} {121102} (\bibinfo {year}
  {2017}{\natexlab{b}})},\ \Eprint {http://arxiv.org/abs/1612.02030}
  {arXiv:1612.02030 [gr-qc]} \BibitemShut {NoStop}%
\bibitem [{\citenamefont {Abbott}\ \emph
  {et~al.}(2019{\natexlab{c}})\citenamefont {Abbott} \emph
  {et~al.}}]{LIGOScientific:2019gaw}%
  \BibitemOpen
  \bibfield  {author} {\bibinfo {author} {\bibfnamefont {B.~P.}\ \bibnamefont
  {Abbott}} \emph {et~al.} (\bibinfo {collaboration} {LIGO, Virgo}),\
  }\bibfield  {title} {\enquote {\bibinfo {title} {{Directional limits on
  persistent gravitational waves using data from Advanced LIGO's first two
  observing runs}},}\ }\href {\doibase 10.1103/PhysRevD.100.062001} {\bibfield
  {journal} {\bibinfo  {journal} {Phys. Rev. D}\ }\textbf {\bibinfo {volume}
  {100}},\ \bibinfo {pages} {062001} (\bibinfo {year} {2019}{\natexlab{c}})},\
  \Eprint {http://arxiv.org/abs/1903.08844} {arXiv:1903.08844 [gr-qc]}
  \BibitemShut {NoStop}%
\bibitem [{\citenamefont {Abbott}\ \emph
  {et~al.}(2021{\natexlab{f}})\citenamefont {Abbott} \emph
  {et~al.}}]{KAGRA:2021mth}%
  \BibitemOpen
  \bibfield  {author} {\bibinfo {author} {\bibfnamefont {R.}~\bibnamefont
  {Abbott}} \emph {et~al.} (\bibinfo {collaboration} {LIGO, Virgo, KAGRA}),\
  }\bibfield  {title} {\enquote {\bibinfo {title} {{Search for anisotropic
  gravitational-wave backgrounds using data from Advanced LIGO and Advanced
  Virgo\textquoteright{}s first three observing runs}},}\ }\href {\doibase
  10.1103/PhysRevD.104.022005} {\bibfield  {journal} {\bibinfo  {journal}
  {Phys. Rev. D}\ }\textbf {\bibinfo {volume} {104}},\ \bibinfo {pages}
  {022005} (\bibinfo {year} {2021}{\natexlab{f}})},\ \Eprint
  {http://arxiv.org/abs/2103.08520} {arXiv:2103.08520 [gr-qc]} \BibitemShut
  {NoStop}%
\bibitem [{\citenamefont {Ade}\ \emph {et~al.}(2016)\citenamefont {Ade} \emph
  {et~al.}}]{Planck:2015fie}%
  \BibitemOpen
  \bibfield  {author} {\bibinfo {author} {\bibfnamefont {P.~A.~R.}\
  \bibnamefont {Ade}} \emph {et~al.} (\bibinfo {collaboration} {Planck}),\
  }\bibfield  {title} {\enquote {\bibinfo {title} {{Planck 2015 results. XIII.
  Cosmological parameters}},}\ }\href {\doibase 10.1051/0004-6361/201525830}
  {\bibfield  {journal} {\bibinfo  {journal} {Astron. Astrophys.}\ }\textbf
  {\bibinfo {volume} {594}},\ \bibinfo {pages} {A13} (\bibinfo {year}
  {2016})},\ \Eprint {http://arxiv.org/abs/1502.01589} {arXiv:1502.01589
  [astro-ph.CO]} \BibitemShut {NoStop}%
\bibitem [{\citenamefont {Phinney}(2001)}]{Phinney:2001di}%
  \BibitemOpen
  \bibfield  {author} {\bibinfo {author} {\bibfnamefont {E.~S.}\ \bibnamefont
  {Phinney}},\ }\bibfield  {title} {\enquote {\bibinfo {title} {{A Practical
  theorem on gravitational wave backgrounds}},}\ }\href@noop {} {\  (\bibinfo
  {year} {2001})},\ \Eprint {http://arxiv.org/abs/astro-ph/0108028}
  {arXiv:astro-ph/0108028} \BibitemShut {NoStop}%
\bibitem [{\citenamefont {Buonanno}\ \emph {et~al.}(2005)\citenamefont
  {Buonanno}, \citenamefont {Sigl}, \citenamefont {Raffelt}, \citenamefont
  {Janka},\ and\ \citenamefont {Muller}}]{Buonanno:2004tp}%
  \BibitemOpen
  \bibfield  {author} {\bibinfo {author} {\bibfnamefont {Alessandra}\
  \bibnamefont {Buonanno}}, \bibinfo {author} {\bibfnamefont {Gunter}\
  \bibnamefont {Sigl}}, \bibinfo {author} {\bibfnamefont {Georg~G.}\
  \bibnamefont {Raffelt}}, \bibinfo {author} {\bibfnamefont {Hans-Thomas}\
  \bibnamefont {Janka}}, \ and\ \bibinfo {author} {\bibfnamefont {Ewald}\
  \bibnamefont {Muller}},\ }\bibfield  {title} {\enquote {\bibinfo {title}
  {{Stochastic gravitational wave background from cosmological supernovae}},}\
  }\href {\doibase 10.1103/PhysRevD.72.084001} {\bibfield  {journal} {\bibinfo
  {journal} {Phys. Rev. D}\ }\textbf {\bibinfo {volume} {72}},\ \bibinfo
  {pages} {084001} (\bibinfo {year} {2005})},\ \Eprint
  {http://arxiv.org/abs/astro-ph/0412277} {arXiv:astro-ph/0412277} \BibitemShut
  {NoStop}%
\bibitem [{\citenamefont {Callister}\ \emph {et~al.}(2016)\citenamefont
  {Callister}, \citenamefont {Sammut}, \citenamefont {Qiu}, \citenamefont
  {Mandel},\ and\ \citenamefont {Thrane}}]{Callister:2016ewt}%
  \BibitemOpen
  \bibfield  {author} {\bibinfo {author} {\bibfnamefont {Thomas}\ \bibnamefont
  {Callister}}, \bibinfo {author} {\bibfnamefont {Letizia}\ \bibnamefont
  {Sammut}}, \bibinfo {author} {\bibfnamefont {Shi}\ \bibnamefont {Qiu}},
  \bibinfo {author} {\bibfnamefont {Ilya}\ \bibnamefont {Mandel}}, \ and\
  \bibinfo {author} {\bibfnamefont {Eric}\ \bibnamefont {Thrane}},\ }\bibfield
  {title} {\enquote {\bibinfo {title} {{The limits of astrophysics with
  gravitational-wave backgrounds}},}\ }\href {\doibase
  10.1103/PhysRevX.6.031018} {\bibfield  {journal} {\bibinfo  {journal} {Phys.
  Rev. X}\ }\textbf {\bibinfo {volume} {6}},\ \bibinfo {pages} {031018}
  (\bibinfo {year} {2016})},\ \Eprint {http://arxiv.org/abs/1604.02513}
  {arXiv:1604.02513 [gr-qc]} \BibitemShut {NoStop}%
\bibitem [{\citenamefont {Cardoso}\ and\ \citenamefont
  {Maselli}(2020)}]{Cardoso:2019rou}%
  \BibitemOpen
  \bibfield  {author} {\bibinfo {author} {\bibfnamefont {Vitor}\ \bibnamefont
  {Cardoso}}\ and\ \bibinfo {author} {\bibfnamefont {Andrea}\ \bibnamefont
  {Maselli}},\ }\bibfield  {title} {\enquote {\bibinfo {title} {{Constraints on
  the astrophysical environment of binaries with gravitational-wave
  observations}},}\ }\href {\doibase 10.1051/0004-6361/202037654} {\bibfield
  {journal} {\bibinfo  {journal} {Astron. Astrophys.}\ }\textbf {\bibinfo
  {volume} {644}},\ \bibinfo {pages} {A147} (\bibinfo {year} {2020})},\ \Eprint
  {http://arxiv.org/abs/1909.05870} {arXiv:1909.05870 [astro-ph.HE]}
  \BibitemShut {NoStop}%
\bibitem [{\citenamefont {Huang}\ \emph {et~al.}(2019)\citenamefont {Huang},
  \citenamefont {Johnson}, \citenamefont {Sagunski}, \citenamefont
  {Sakellariadou},\ and\ \citenamefont {Zhang}}]{Huang:2018pbu}%
  \BibitemOpen
  \bibfield  {author} {\bibinfo {author} {\bibfnamefont {Junwu}\ \bibnamefont
  {Huang}}, \bibinfo {author} {\bibfnamefont {Matthew~C.}\ \bibnamefont
  {Johnson}}, \bibinfo {author} {\bibfnamefont {Laura}\ \bibnamefont
  {Sagunski}}, \bibinfo {author} {\bibfnamefont {Mairi}\ \bibnamefont
  {Sakellariadou}}, \ and\ \bibinfo {author} {\bibfnamefont {Jun}\ \bibnamefont
  {Zhang}},\ }\bibfield  {title} {\enquote {\bibinfo {title} {{Prospects for
  axion searches with Advanced LIGO through binary mergers}},}\ }\href
  {\doibase 10.1103/PhysRevD.99.063013} {\bibfield  {journal} {\bibinfo
  {journal} {Phys. Rev. D}\ }\textbf {\bibinfo {volume} {99}},\ \bibinfo
  {pages} {063013} (\bibinfo {year} {2019})},\ \Eprint
  {http://arxiv.org/abs/1807.02133} {arXiv:1807.02133 [hep-ph]} \BibitemShut
  {NoStop}%
\bibitem [{\citenamefont {Murray}\ and\ \citenamefont
  {Dermott}(2000)}]{Murray:2000ssd}%
  \BibitemOpen
  \bibfield  {author} {\bibinfo {author} {\bibfnamefont {C.~D.}\ \bibnamefont
  {Murray}}\ and\ \bibinfo {author} {\bibfnamefont {S.~F.}\ \bibnamefont
  {Dermott}},\ }\href
  {http://www.cambridge.org/us/catalogue/catalogue.asp?isbn=0521575974.~Cambridge}
  {\emph {\bibinfo {title} {{Solar System Dynamics}}}}\ (\bibinfo  {publisher}
  {Cambridge University Press},\ \bibinfo {year} {2000})\BibitemShut {NoStop}%
\bibitem [{\citenamefont {Robitaille}\ \emph {et~al.}(2013)\citenamefont
  {Robitaille} \emph {et~al.}}]{Astropy:2013muo}%
  \BibitemOpen
  \bibfield  {author} {\bibinfo {author} {\bibfnamefont {Thomas~P.}\
  \bibnamefont {Robitaille}} \emph {et~al.} (\bibinfo {collaboration}
  {Astropy}),\ }\bibfield  {title} {\enquote {\bibinfo {title} {{Astropy: A
  Community Python Package for Astronomy}},}\ }\href {\doibase
  10.1051/0004-6361/201322068} {\bibfield  {journal} {\bibinfo  {journal}
  {Astron. Astrophys.}\ }\textbf {\bibinfo {volume} {558}},\ \bibinfo {pages}
  {A33} (\bibinfo {year} {2013})},\ \Eprint {http://arxiv.org/abs/1307.6212}
  {arXiv:1307.6212 [astro-ph.IM]} \BibitemShut {NoStop}%
\bibitem [{\citenamefont {Price-Whelan}\ \emph {et~al.}(2018)\citenamefont
  {Price-Whelan} \emph {et~al.}}]{Astropy:2018wqo}%
  \BibitemOpen
  \bibfield  {author} {\bibinfo {author} {\bibfnamefont {A.~M.}\ \bibnamefont
  {Price-Whelan}} \emph {et~al.} (\bibinfo {collaboration} {Astropy}),\
  }\bibfield  {title} {\enquote {\bibinfo {title} {{The Astropy Project:
  Building an Open-science Project and Status of the v2.0 Core Package}},}\
  }\href {\doibase 10.3847/1538-3881/aabc4f} {\bibfield  {journal} {\bibinfo
  {journal} {Astron. J.}\ }\textbf {\bibinfo {volume} {156}},\ \bibinfo {pages}
  {123} (\bibinfo {year} {2018})},\ \Eprint {http://arxiv.org/abs/1801.02634}
  {arXiv:1801.02634} \BibitemShut {NoStop}%
\bibitem [{\citenamefont {Allen}\ and\ \citenamefont
  {Romano}(1999)}]{Allen:1997ad}%
  \BibitemOpen
  \bibfield  {author} {\bibinfo {author} {\bibfnamefont {Bruce}\ \bibnamefont
  {Allen}}\ and\ \bibinfo {author} {\bibfnamefont {Joseph~D.}\ \bibnamefont
  {Romano}},\ }\bibfield  {title} {\enquote {\bibinfo {title} {{Detecting a
  stochastic background of gravitational radiation: Signal processing
  strategies and sensitivities}},}\ }\href {\doibase
  10.1103/PhysRevD.59.102001} {\bibfield  {journal} {\bibinfo  {journal} {Phys.
  Rev. D}\ }\textbf {\bibinfo {volume} {59}},\ \bibinfo {pages} {102001}
  (\bibinfo {year} {1999})},\ \Eprint {http://arxiv.org/abs/gr-qc/9710117}
  {arXiv:gr-qc/9710117} \BibitemShut {NoStop}%
\bibitem [{\citenamefont {Meyers}\ \emph {et~al.}(2020)\citenamefont {Meyers},
  \citenamefont {Martinovic}, \citenamefont {Christensen},\ and\ \citenamefont
  {Sakellariadou}}]{Meyers:2020qrb}%
  \BibitemOpen
  \bibfield  {author} {\bibinfo {author} {\bibfnamefont {Patrick~M.}\
  \bibnamefont {Meyers}}, \bibinfo {author} {\bibfnamefont {Katarina}\
  \bibnamefont {Martinovic}}, \bibinfo {author} {\bibfnamefont {Nelson}\
  \bibnamefont {Christensen}}, \ and\ \bibinfo {author} {\bibfnamefont {Mairi}\
  \bibnamefont {Sakellariadou}},\ }\bibfield  {title} {\enquote {\bibinfo
  {title} {{Detecting a stochastic gravitational-wave background in the
  presence of correlated magnetic noise}},}\ }\href {\doibase
  10.1103/PhysRevD.102.102005} {\bibfield  {journal} {\bibinfo  {journal}
  {Phys. Rev. D}\ }\textbf {\bibinfo {volume} {102}},\ \bibinfo {pages}
  {102005} (\bibinfo {year} {2020})},\ \Eprint
  {http://arxiv.org/abs/2008.00789} {arXiv:2008.00789 [gr-qc]} \BibitemShut
  {NoStop}%
\bibitem [{\citenamefont {Abbott}\ \emph
  {et~al.}(2021{\natexlab{g}})\citenamefont {Abbott} \emph {et~al.}}]{O3_SHD}%
  \BibitemOpen
  \bibfield  {author} {\bibinfo {author} {\bibfnamefont {R.}~\bibnamefont
  {Abbott}} \emph {et~al.} (\bibinfo {collaboration} {LIGO, Virgo, KAGRA}),\
  }\href@noop {} {\enquote {\bibinfo {title} {{Data products and supplemental
  information for O3 stochastic directional paper}},}\ }\bibinfo {howpublished}
  {\url{https://dcc.ligo.org/LIGO-G2002165/public}} (\bibinfo {year}
  {2021}{\natexlab{g}})\BibitemShut {NoStop}%
\bibitem [{\citenamefont {Abbott}\ \emph {et~al.}(2004)\citenamefont {Abbott}
  \emph {et~al.}}]{LIGOScientific:2003jxj}%
  \BibitemOpen
  \bibfield  {author} {\bibinfo {author} {\bibfnamefont {B.}~\bibnamefont
  {Abbott}} \emph {et~al.} (\bibinfo {collaboration} {LIGO}),\ }\bibfield
  {title} {\enquote {\bibinfo {title} {{Analysis of first LIGO science data for
  stochastic gravitational waves}},}\ }\href {\doibase
  10.1103/PhysRevD.69.122004} {\bibfield  {journal} {\bibinfo  {journal} {Phys.
  Rev. D}\ }\textbf {\bibinfo {volume} {69}},\ \bibinfo {pages} {122004}
  (\bibinfo {year} {2004})},\ \Eprint {http://arxiv.org/abs/gr-qc/0312088}
  {arXiv:gr-qc/0312088} \BibitemShut {NoStop}%
\bibitem [{\citenamefont {Lazzarini}(2004)}]{bias_factor}%
  \BibitemOpen
  \bibfield  {author} {\bibinfo {author} {\bibfnamefont {Albert}\ \bibnamefont
  {Lazzarini}} (\bibinfo {collaboration} {LIGO}),\ }\href@noop {} {\enquote
  {\bibinfo {title} {Bias from power spectrum measurement in parameter
  estimation for the stochastic gravitational wave background},}\ }\bibinfo
  {howpublished} {\url{https://dcc.ligo.org/LIGO-T040128/public}} (\bibinfo
  {year} {2004})\BibitemShut {NoStop}%
\bibitem [{\citenamefont {Lazzarini}\ and\ \citenamefont
  {Romano}(2004)}]{Overlapping_windows}%
  \BibitemOpen
  \bibfield  {author} {\bibinfo {author} {\bibfnamefont {Albert}\ \bibnamefont
  {Lazzarini}}\ and\ \bibinfo {author} {\bibfnamefont {Joe}\ \bibnamefont
  {Romano}} (\bibinfo {collaboration} {LIGO}),\ }\href@noop {} {\enquote
  {\bibinfo {title} {Use of overlapping windows in the stochastic background
  search},}\ }\bibinfo {howpublished}
  {\url{https://dcc.ligo.org/T040089/public}} (\bibinfo {year}
  {2004})\BibitemShut {NoStop}%
\bibitem [{\citenamefont {Abbott}\ \emph
  {et~al.}(2019{\natexlab{d}})\citenamefont {Abbott} \emph
  {et~al.}}]{O2_directional_notch_list}%
  \BibitemOpen
  \bibfield  {author} {\bibinfo {author} {\bibfnamefont {R.}~\bibnamefont
  {Abbott}} \emph {et~al.} (\bibinfo {collaboration} {LIGO}),\ }\href@noop {}
  {\enquote {\bibinfo {title} {{Data for `Search for the isotropic stochastic
  background using data from Advanced LIGO\textquoteright{}s second observing
  run'}},}\ }\bibinfo {howpublished}
  {\url{https://dcc.ligo.org/T1900058/public}} (\bibinfo {year}
  {2019}{\natexlab{d}})\BibitemShut {NoStop}%
\bibitem [{\citenamefont {Abbott}\ \emph {et~al.}(2016)\citenamefont {Abbott}
  \emph {et~al.}}]{LIGOScientific:2016gtq}%
  \BibitemOpen
  \bibfield  {author} {\bibinfo {author} {\bibfnamefont {B.~P.}\ \bibnamefont
  {Abbott}} \emph {et~al.} (\bibinfo {collaboration} {LIGO, Virgo}),\
  }\bibfield  {title} {\enquote {\bibinfo {title} {{Characterization of
  transient noise in Advanced LIGO relevant to gravitational wave signal
  GW150914}},}\ }\href {\doibase 10.1088/0264-9381/33/13/134001} {\bibfield
  {journal} {\bibinfo  {journal} {Class. Quant. Grav.}\ }\textbf {\bibinfo
  {volume} {33}},\ \bibinfo {pages} {134001} (\bibinfo {year} {2016})},\
  \Eprint {http://arxiv.org/abs/1602.03844} {arXiv:1602.03844 [gr-qc]}
  \BibitemShut {NoStop}%
\bibitem [{\citenamefont {Thrane}\ \emph {et~al.}(2013)\citenamefont {Thrane},
  \citenamefont {Christensen},\ and\ \citenamefont
  {Schofield}}]{Thrane:2013npa}%
  \BibitemOpen
  \bibfield  {author} {\bibinfo {author} {\bibfnamefont {Eric}\ \bibnamefont
  {Thrane}}, \bibinfo {author} {\bibfnamefont {Nelson}\ \bibnamefont
  {Christensen}}, \ and\ \bibinfo {author} {\bibfnamefont {Robert}\
  \bibnamefont {Schofield}},\ }\bibfield  {title} {\enquote {\bibinfo {title}
  {{Correlated magnetic noise in global networks of gravitational-wave
  interferometers: observations and implications}},}\ }\href {\doibase
  10.1103/PhysRevD.87.123009} {\bibfield  {journal} {\bibinfo  {journal} {Phys.
  Rev. D}\ }\textbf {\bibinfo {volume} {87}},\ \bibinfo {pages} {123009}
  (\bibinfo {year} {2013})},\ \Eprint {http://arxiv.org/abs/1303.2613}
  {arXiv:1303.2613 [astro-ph.IM]} \BibitemShut {NoStop}%
\bibitem [{\citenamefont {Thrane}\ \emph {et~al.}(2014)\citenamefont {Thrane},
  \citenamefont {Christensen}, \citenamefont {Schofield},\ and\ \citenamefont
  {Effler}}]{Thrane:2014yza}%
  \BibitemOpen
  \bibfield  {author} {\bibinfo {author} {\bibfnamefont {E.}~\bibnamefont
  {Thrane}}, \bibinfo {author} {\bibfnamefont {N.}~\bibnamefont {Christensen}},
  \bibinfo {author} {\bibfnamefont {R.~M.~S.}\ \bibnamefont {Schofield}}, \
  and\ \bibinfo {author} {\bibfnamefont {A.}~\bibnamefont {Effler}},\
  }\bibfield  {title} {\enquote {\bibinfo {title} {{Correlated noise in
  networks of gravitational-wave detectors: subtraction and mitigation}},}\
  }\href {\doibase 10.1103/PhysRevD.90.023013} {\bibfield  {journal} {\bibinfo
  {journal} {Phys. Rev. D}\ }\textbf {\bibinfo {volume} {90}},\ \bibinfo
  {pages} {023013} (\bibinfo {year} {2014})},\ \Eprint
  {http://arxiv.org/abs/1406.2367} {arXiv:1406.2367 [astro-ph.IM]} \BibitemShut
  {NoStop}%
\bibitem [{\citenamefont {Biwer}\ \emph {et~al.}(2016)\citenamefont {Biwer}
  \emph {et~al.}}]{LIGO_data_cut_04}%
  \BibitemOpen
  \bibfield  {author} {\bibinfo {author} {\bibfnamefont {C.}~\bibnamefont
  {Biwer}} \emph {et~al.} (\bibinfo {collaboration} {LIGO}),\ }\href@noop {}
  {\enquote {\bibinfo {title} {Validating gravitational-wave detections: The
  advanced ligo hardware injection system},}\ }\bibinfo {howpublished}
  {\url{https://dcc.ligo.org/LIGO-P1600285/public}} (\bibinfo {year}
  {2016})\BibitemShut {NoStop}%
\bibitem [{\citenamefont {Punturo}\ \emph {et~al.}(2010)\citenamefont {Punturo}
  \emph {et~al.}}]{Punturo:2010zz}%
  \BibitemOpen
  \bibfield  {author} {\bibinfo {author} {\bibfnamefont {M.}~\bibnamefont
  {Punturo}} \emph {et~al.},\ }\bibfield  {title} {\enquote {\bibinfo {title}
  {{The Einstein Telescope: A third-generation gravitational wave
  observatory}},}\ }\href {\doibase 10.1088/0264-9381/27/19/194002} {\bibfield
  {journal} {\bibinfo  {journal} {Class. Quant. Grav.}\ }\textbf {\bibinfo
  {volume} {27}},\ \bibinfo {pages} {194002} (\bibinfo {year}
  {2010})}\BibitemShut {NoStop}%
\bibitem [{\citenamefont {Reitze}\ \emph {et~al.}(2019)\citenamefont {Reitze}
  \emph {et~al.}}]{Reitze:2019iox}%
  \BibitemOpen
  \bibfield  {author} {\bibinfo {author} {\bibfnamefont {David}\ \bibnamefont
  {Reitze}} \emph {et~al.},\ }\bibfield  {title} {\enquote {\bibinfo {title}
  {{Cosmic Explorer: The U.S. Contribution to Gravitational-Wave Astronomy
  beyond LIGO}},}\ }\href {https://baas.aas.org/pub/2020n7i035} {\bibfield
  {journal} {\bibinfo  {journal} {Bull. Am. Astron. Soc.}\ }\textbf {\bibinfo
  {volume} {51}},\ \bibinfo {pages} {035} (\bibinfo {year} {2019})},\ \Eprint
  {http://arxiv.org/abs/1907.04833} {arXiv:1907.04833 [astro-ph.IM]}
  \BibitemShut {NoStop}%
\bibitem [{\citenamefont {design team}(2018)}]{ETcurve}%
  \BibitemOpen
  \bibfield  {author} {\bibinfo {author} {\bibfnamefont {ET}~\bibnamefont
  {design team}},\ }\href@noop {} {\enquote {\bibinfo {title} {{ET-D
  sensitivity curve}},}\ }\bibinfo {howpublished}
  {\url{https://apps.et-gw.eu/tds/?content=3&r=14065}} (\bibinfo {year}
  {2018})\BibitemShut {NoStop}%
\bibitem [{\citenamefont {Talukder}\ \emph {et~al.}(2014)\citenamefont
  {Talukder}, \citenamefont {Thrane}, \citenamefont {Bose},\ and\ \citenamefont
  {Regimbau}}]{Talukder:2014eba}%
  \BibitemOpen
  \bibfield  {author} {\bibinfo {author} {\bibfnamefont {Dipongkar}\
  \bibnamefont {Talukder}}, \bibinfo {author} {\bibfnamefont {Eric}\
  \bibnamefont {Thrane}}, \bibinfo {author} {\bibfnamefont {Sukanta}\
  \bibnamefont {Bose}}, \ and\ \bibinfo {author} {\bibfnamefont {Tania}\
  \bibnamefont {Regimbau}},\ }\bibfield  {title} {\enquote {\bibinfo {title}
  {{Measuring neutron-star ellipticity with measurements of the stochastic
  gravitational-wave background}},}\ }\href {\doibase
  10.1103/PhysRevD.89.123008} {\bibfield  {journal} {\bibinfo  {journal} {Phys.
  Rev. D}\ }\textbf {\bibinfo {volume} {89}},\ \bibinfo {pages} {123008}
  (\bibinfo {year} {2014})},\ \Eprint {http://arxiv.org/abs/1404.4025}
  {arXiv:1404.4025 [gr-qc]} \BibitemShut {NoStop}%
\bibitem [{\citenamefont {Agarwal}\ \emph {et~al.}(2022)\citenamefont
  {Agarwal}, \citenamefont {Suresh}, \citenamefont {Mandic}, \citenamefont
  {Matas},\ and\ \citenamefont {Regimbau}}]{Agarwal:2022lvk}%
  \BibitemOpen
  \bibfield  {author} {\bibinfo {author} {\bibfnamefont {Deepali}\ \bibnamefont
  {Agarwal}}, \bibinfo {author} {\bibfnamefont {Jishnu}\ \bibnamefont
  {Suresh}}, \bibinfo {author} {\bibfnamefont {Vuk}\ \bibnamefont {Mandic}},
  \bibinfo {author} {\bibfnamefont {Andrew}\ \bibnamefont {Matas}}, \ and\
  \bibinfo {author} {\bibfnamefont {Tania}\ \bibnamefont {Regimbau}},\
  }\bibfield  {title} {\enquote {\bibinfo {title} {{Targeted search for the
  stochastic gravitational-wave background from the galactic millisecond pulsar
  population}},}\ }\href@noop {} {\  (\bibinfo {year} {2022})},\ \Eprint
  {http://arxiv.org/abs/2204.08378} {arXiv:2204.08378 [gr-qc]} \BibitemShut
  {NoStop}%
\bibitem [{\citenamefont {Zhu}\ \emph {et~al.}(2013)\citenamefont {Zhu},
  \citenamefont {Howell}, \citenamefont {Blair},\ and\ \citenamefont
  {Zhu}}]{Zhu:2012xw}%
  \BibitemOpen
  \bibfield  {author} {\bibinfo {author} {\bibfnamefont {Xing-Jiang}\
  \bibnamefont {Zhu}}, \bibinfo {author} {\bibfnamefont {Eric~J.}\ \bibnamefont
  {Howell}}, \bibinfo {author} {\bibfnamefont {David~G.}\ \bibnamefont
  {Blair}}, \ and\ \bibinfo {author} {\bibfnamefont {Zong-Hong}\ \bibnamefont
  {Zhu}},\ }\bibfield  {title} {\enquote {\bibinfo {title} {{On the
  gravitational wave background from compact binary coalescences in the band of
  ground-based interferometers}},}\ }\href {\doibase 10.1093/mnras/stt207}
  {\bibfield  {journal} {\bibinfo  {journal} {Mon. Not. Roy. Astron. Soc.}\
  }\textbf {\bibinfo {volume} {431}},\ \bibinfo {pages} {882--899} (\bibinfo
  {year} {2013})},\ \Eprint {http://arxiv.org/abs/1209.0595} {arXiv:1209.0595
  [gr-qc]} \BibitemShut {NoStop}%
\bibitem [{\citenamefont {Regimbau}\ \emph {et~al.}(2017)\citenamefont
  {Regimbau}, \citenamefont {Evans}, \citenamefont {Christensen}, \citenamefont
  {Katsavounidis}, \citenamefont {Sathyaprakash},\ and\ \citenamefont
  {Vitale}}]{Regimbau:2016ike}%
  \BibitemOpen
  \bibfield  {author} {\bibinfo {author} {\bibfnamefont {T.}~\bibnamefont
  {Regimbau}}, \bibinfo {author} {\bibfnamefont {M.}~\bibnamefont {Evans}},
  \bibinfo {author} {\bibfnamefont {N.}~\bibnamefont {Christensen}}, \bibinfo
  {author} {\bibfnamefont {E.}~\bibnamefont {Katsavounidis}}, \bibinfo {author}
  {\bibfnamefont {B.}~\bibnamefont {Sathyaprakash}}, \ and\ \bibinfo {author}
  {\bibfnamefont {S.}~\bibnamefont {Vitale}},\ }\bibfield  {title} {\enquote
  {\bibinfo {title} {{Digging deeper: Observing primordial gravitational waves
  below the binary black hole produced stochastic background}},}\ }\href
  {\doibase 10.1103/PhysRevLett.118.151105} {\bibfield  {journal} {\bibinfo
  {journal} {Phys. Rev. Lett.}\ }\textbf {\bibinfo {volume} {118}},\ \bibinfo
  {pages} {151105} (\bibinfo {year} {2017})},\ \Eprint
  {http://arxiv.org/abs/1611.08943} {arXiv:1611.08943 [astro-ph.CO]}
  \BibitemShut {NoStop}%
\end{thebibliography}%
\end{document}